\pdfoutput=1
\documentclass[a4paper,fleqn,usenatbib,useAMS]{mnras}

\usepackage{graphicx}	% Including figure files
\usepackage{amsmath}	% Advanced maths commands
\usepackage{amssymb}	% Extra maths symbols
\usepackage{multicol}        % Multi-column entries in tables
\usepackage{bm}		% Bold maths symbols, including upright Greek
\usepackage{pdflscape}	% Landscape pages

\usepackage{subfigure}
\usepackage{float}
\usepackage{graphicx}
\usepackage{booktabs}
\usepackage{longtable}
\newcommand{\kph}{km\,s$^{-1}$\,}

\newcommand{\angstrom}{${\rm \AA}$}

\newcommand{\msol}{$\rm {M_{\odot}}$}
\newcommand{\msolyr}{$\rm{M_{\odot}}$\,yr$^{-1}$}

\newcommand{\Sigmas}{${\Sigma}_{\star}$\,}

\newcommand{\sigmao}{$\sigma_{\rm{los}}$\,}
\newcommand{\sigmam}{$\sigma_{\rm{mod}}$\,}
\newcommand{\sigmaR}{$\sigma_{R}$\,}

\newcommand{\sigmaRR}{$\sigma_{R}(R)$\,}
\newcommand{\sigmaz}{$\sigma_{z}$\,}
\newcommand{\sigmap}{$\sigma_{\phi}$\,}

\newcommand{\medsigmaR}{$\langle\,\sigma_{R}\,\rangle$}
\newcommand{\medsigmaRfit}{$\langle\,\sigma_{R}\,\rangle _{\rm{fit} }$}
\newcommand{\medsigmalos}{$\langle\,\sigma_{\rm{los}}\,\rangle$}
\newcommand{\medsigmam}{$\langle\,\sigma_{\rm mod}\,\rangle$}
\newcommand{\medsigmaRobs}{$\langle\,\sigma_{R}\,\rangle _{\rm{obs}}$}
\newcommand{\medsigmaRmod}{$\langle\,\sigma_{R}\,\rangle _{\rm{mod}}$}

\newcommand{\gml}{$\sqrt {GM_{\star}/l_{\rm \star}} $}

\usepackage{xcolor}
\usepackage{soul}

\def\apjl{ApJL}
\def\aj{AJ}
\def\apj{ApJ}

\def\aap{A\&A}
\def\mnras{MNRAS}

\defcitealias{fal17}{F-B17}
\defcitealias{kal17b}{K17}
\defcitealias{bol17}{B17}
\defcitealias{ler08}{L08}

\usepackage[T1]{fontenc}
\usepackage{ae,aecompl}

\usepackage{mathptmx}
\title[The stellar velocity dispersion in nearby spirals]{The stellar velocity dispersion in nearby spirals: radial profiles and correlations}
\author[K. M. Mogotsi and A. B. Romeo ]{Keoikantse Moses Mogotsi $^{1,2}$\thanks{E-mail:
moses.mog@gmail.com} and
Alessandro B. Romeo$^{3}$ \\
$^{1}$South African Astronomical Observatory, P.O. Box 9, Observatory, Cape Town, South Africa \\
$^{2}$Southern African Large Telescope,  P.O. Box 9, Observatory, Cape Town, South Africa\\
$^{3}$Department of Space, Earth and Environment, Chalmers University of Technology, SE-41296 Gothenburg, Sweden   \\
}

\date{}

\pubyear{}

\begin{document}
\label{firstpage}
\pagerange{\pageref{firstpage}--\pageref{lastpage}}
\maketitle

\begin{abstract}
The stellar velocity dispersion, $\sigma$, is a quantity of crucial importance for spiral galaxies, where it enters fundamental dynamical  processes such as gravitational instability and disc heating.  Here we analyse a sample of 34 nearby spirals from the Calar Alto Legacy Integral Field Area (CALIFA) spectroscopic survey, deproject the line-of-sight $\sigma$ to $\sigma_{R}$ and present reliable radial profiles of $\sigma_{R}$ as well as accurate measurements of $\langle\sigma_{R}\rangle$, the radial average of $\sigma_{R}$ over one effective (half-light) radius.  We show that there is a trend for $\sigma_{R}$ to increase with decreasing $R$, that $\langle\sigma_{R}\rangle$ correlates with stellar mass ($M_{\star}$) and tested correlations with other galaxy properties.  The most significant and strongest correlation is the one with $M_{\star}$: $\langle\sigma_{R}\rangle \propto M_{\star}^{0.5}$.  This tight scaling relation is applicable to spiral galaxies of type Sa--Sd and stellar mass $M_{\star}\approx10^{9.5}\mbox{--}10^{11.5}\ \mbox{M}_{\odot}$.  Simple models that relate $\sigma_{R}$ to the stellar surface density and disc scale length roughly reproduce that scaling, but overestimate $\langle\sigma_{R}\rangle$ significantly. 
\end{abstract}


\section{INTRODUCTION}

The stellar velocity dispersion is an important parameter in stellar disc dynamics and has a wide range of applications.  The various velocity dispersion components are used to study the distribution of stars near the solar neighbourhood \citep[e.g.,][]{deh98,dehbin98,tia15} and how stars of different ages are distributed \citep[e.g.,][]{wie77,dehbin98,bin00}.  This is used to make more detailed characterization of the structure and evolution of the Milky Way's stellar disc and its different components.  These detailed local observations show the anisotropy between the radial, azimuthal and vertical stellar velocity dispersion components such that \sigmaR $>$ \sigmap $>$ \sigmaz.  The ratios of these components (anisotropy parameters) are often thought of as the velocity ellipsoid \citep[e.g.,][]{sch07} and are crucial to quantifying the anisotropy and understanding its causes \citep*[e.g.,][]{spi51,jen90, sha03, ger12, pin18}.  In particular, \sigmaz/\sigmaR has a minimum of $0.3$ due to the bending instability \citep[][]{rod13} and is used to constrain these ``disc heating'' processes.  \sigmaz is used to measure the mass-to-light-ratio of galactic discs \citep[e.g.,][]{vdkru81, vdkru88, ber10,an18} .  In kinematic studies, \sigmap/\sigmaR is used to check the validity of the epicyclic approximation for stellar motions in the plane of a disc and \sigmaR is used to correct rotation curves for asymmetric drift \citep[e.g.,][]{bin08}.         

The stellar radial velocity dispersion, $\sigma_{R}$, is also one of the quantities that most radically affect the onset of gravitational instabilities in galaxy discs.  It enters \citeauthor{tom64}'s~(\citeyear{tom64}) stability criterion $Q\equiv\kappa\sigma_{R}/(3.36\,G\Sigma)\geq1$ for infinitesimally thin stellar discs, as well as in more modern and advanced local stability analyses for multi-component \citep[e.g.,][]{raf01,ler08,wes14} and realistically thick \citep[e.g.,][]{rom13} discs.  \cite{rom17} showed that stars, and not molecular or atomic gas, are the primary driver of disc instabilities in spiral galaxies, at least at the spatial resolution of current extragalactic surveys.  This is true even for a powerful starburst and Seyfert galaxy like NGC 1068 \citep[][]{rom16}.  Thus $\sigma_{R}$ is now recognized, more confidently than before, as a crucial quantity for disc instability.

It is difficult to obtain accurate and resolved measurements of stellar velocity dispersions for a large sample of galaxies and velocity dispersion components are difficult to disentangle from line-of-sight measurements \citep[e.g.,][]{ger97,ger00,sha03,ger12,che18,pin18}.  This is why disc stability analyses use model-based estimates of \sigmaR and make assumptions about the anisotropy parameters \citep[e.g.,][]{ler08,rom17}. 

The advent of integral field surveys such as SAMI \mbox{\citep[][]{all15}} and MaNGA \mbox{\citep[][]{bun15}} is increasing the number of galaxies with measured stellar kinematics.  The Calar Alto Legacy Integral Field Area (CALIFA) survey \citep[][]{san12} is a spatially resolved IFU spectroscopic survey of $\sim 600$ nearby galaxies.  The survey provides unprecedented detailed stellar kinematics for such a large and diverse sample of galaxies \citep[e.g.,][]{san17,fal17,kal17b}.  This enables a detailed study of stellar velocity dispersions out to one effective radius and to test stellar dispersion dispersion models.  {\emph{Therefore we aim to use this wealth of quality data to calculate \sigmaR.  We follow this by studying the radial behaviour of \sigmaR, its relation to galaxy properties and to test stellar velocity dispersion models for a sample of spiral galaxies across the Hubble sequence}}. 

We organize the paper as follows. The data is described in Sect. 2, the method and details about calculation of the \sigmaR and model-based dispersions is in Sect. 3\,.  The results of the radial analysis, comparisons between observed and model-based dispersions and relation to galaxy parameters are described in Sect. 4\,. These results are discussed in Sect. 5\, and conclusions are in Sect. 6.

\section{GALAXY SAMPLE AND DATA}

This study is based on a sample of 34 nearby ($D < 122$\,Mpc) spiral galaxies from the CALIFA survey \citep[][]{san12}.  The sample consists of Sa to Sd galaxies for which resolved stellar velocity dispersions, accurate stellar circular-speed curves, molecular gas data, star formation rates, stellar masses and stellar scale lengths are all publicly available.  These are the data needed to calculate stellar radial velocity dispersions and test their correlations with galaxy properties, which we study in this paper and the following ones.  The source of line-of-sight stellar velocity dispersions is the CALIFA high resolution observations (using the V1200 grating to achieve R $\sim 1650$ at a wavelength of $\sim 4500$\,\angstrom) by \citet[][hereafter \citetalias{fal17}]{fal17}, with a velocity resolution of $\sigma \sim 72$\,\kph\,.  We obtain molecular gas data from the EDGE-CALIFA, survey which is a resolved CO follow-up survey of 126 CALIFA galaxies with the CARMA interferometer by \citet[][hereafter \citetalias{bol17}]{bol17}. It has yielded good quality molecular gas data used in studies of the molecular gas properties of galaxies and in the role of gas and star formation in galaxy evolution.  Finally we obtain stellar circular-speed curves and dispersion anisotropy parameters from the study  \citet[][hereafter \citetalias{kal17b}]{kal17b}, who use the axisymmetric Jeans anisotropic multi-Gaussian expansion dynamical method \citep[][]{cap08} to derive these values.  Only 34 galaxies in the CALIFA sample have the requisite publicly available data at high enough quality for this analysis.  The data requirements, sources of data and samples of galaxies with the relevant publicly available data are summarized in \mbox{Table \ref{tab:data}} .

We also select a subsample of galaxies for which stellar surface density data are available.  This subsample consists of 24 galaxies and is crucial to compare the trends between \sigmaR and the modeled velocity dispersion \sigmam across a wide range of galaxy morphologies.  \citetalias{bol17} also use surface density maps to determine the exponential scale lengths of the galaxies which were used in this analysis.  We obtain the stellar surface density \Sigmas maps from \citet[][]{san16}, who developed a pipeline called P{\sc{ipe}}3D to determine dust-corrected \Sigmas of CALIFA galaxies from the low resolution CALIFA Data Release 2 \citep[][]{san12,wal14,gar15} V500 observations using stellar population fitting.  It should be noted that the stellar masses for the entire sample were taken from \mbox{\citetalias{bol17}}, these values are the summation of stellar surface density maps determined using P{\sc{ipe}}3D but they only publicly provide the stellar masses for these galaxies, hence still limiting our surface density subsample to 24 galaxies.   

The maps and data used in this analysis are derived from Voronoi 2D binned \citep[][]{cap03} data cubes.  The galaxy sample covers a wide range of properties such as Hubble types ranging between Sa and Sd, stellar masses ranging between $9.84$ and $11.27$ $\rm{log(M_{\star}/M_{\odot})}$ and star formation rates between $0.7$ and $15.1$\,\msolyr\,.  The global properties of the galaxy sample are shown in Table \ref{tab:sampinfo}.  We use the galaxy properties, dispersion maps, stellar surface density maps, circular-speed curves and dispersion anisotropy values for our analysis.

\begin{table}
\caption{Sample sizes of galaxies with relevant data.}
\label{tab:data}
\begin{tabular}{|l|l|}

\hline 
\multicolumn{1}{l}{Data} & \multicolumn{1}{l}{N} \\
\hline

\sigmao, CSC$^{\rm a}$, $M_{\star}$   & $74$  \tabularnewline
\sigmao, CSC$^{\rm a}$, $M_{\star}$, $M_{\rm{mol}}$,$\rm{SFR}$, $l_{\star}$    & $34^{\rm b}$  \tabularnewline
\sigmao, CSC$^{\rm a}$, $M_{\star}$, $M_{\rm{mol}}$,$\rm{SFR}$, $l_{\star}$, \Sigmas   & $24$  \tabularnewline

\hline 

{\bf{Notes.}} \tabularnewline
\multicolumn{2}{l}{Column 1: Data; }\\
 \multicolumn{2}{l}{Column 2: Number of Sa-Sd galaxies with}\\ 
 \multicolumn{2}{l}{relevant publicaly available data.  } \\
 \multicolumn{2}{l}{Sources of data: \sigmao from \citetalias{fal17}; CSC$^{\rm a}$ } \\
 %\multicolumn{2}{l}{CSC$^{\rm a}$ from \citetalias{kal17b};} \\
 \multicolumn{2}{l}{from \citetalias{kal17b}; $M_{\star}$, $M_{\rm{mol}}$,$\rm{SFR}$ and $l_{\star}$ from   } \\
 \multicolumn{2}{l}{\citetalias{bol17}; and \Sigmas from CALIFA DR2 database;    } \\
 %\multicolumn{2}{l}{\sigmao[\citetalias{fal17}],CSC$^{\rm a}$[\citetalias{kal17b}], $M_{\star}$,$M_{\rm{mol}}$,[\citetalias{bol17}],   } \\
 %\multicolumn{2}{l}{\Sigmas:    } \\
 \multicolumn{2}{l}{$^{\rm a}$ Circular-speed curve  } \\
 \multicolumn{2}{l}{$^{\rm b}$ The 34 galaxies includes NGC2730 which  }\\
 \multicolumn{2}{l}{has $l_{\star}$ calculated from $r_{\rm{e}}$.  } \\

\end{tabular}
%\tablefoot{Sample properties}
\end{table}

\begin{table*}
\centering
\caption{Galaxy properties.}
\label{tab:sampinfo}
\begin{tabular}{llrrrrrrrr}

\hline 
\multicolumn{1}{l}{Name} & \multicolumn{1}{l}{Type} & \multicolumn{1}{c}{$\sigma_{z}/\sigma_{R}$} & \multicolumn{1}{c}{12$\rm{+\log(O/H)}$} &\multicolumn{1}{c}{$\rm{log}$\,$M_{\star}$} & \multicolumn{1}{c}{$\rm{log}$\,$M_{\rm{mol}}$} & \multicolumn{1}{c}{$\rm{log\,SFR}$} &\multicolumn{1}{c}{$l_{\star}$} & \multicolumn{1}{c}{$l_{\rm{mol}}$} & \multicolumn{1}{c}{$l_{\rm{SFR}}$} \\

\multicolumn{1}{l}{} & \multicolumn{1}{l}{}  & \multicolumn{1}{c}{}  & \multicolumn{1}{c}{}  & \multicolumn{1}{c}{[\msol]} & \multicolumn{1}{c}{[\msol]} & \multicolumn{1}{c}{[\msolyr]} &\multicolumn{1}{c}{[kpc]} & \multicolumn{1}{c}{[kpc]} & \multicolumn{1}{c}{[kpc]}  \\

%\hline
\multicolumn{1}{l}{(1)} & \multicolumn{1}{c}{(2)} & \multicolumn{1}{c}{(3)} & \multicolumn{1}{c}{(4)} & \multicolumn{1}{c}{(5)} & \multicolumn{1}{c}{(6)} & \multicolumn{1}{c}{(7)} & \multicolumn{1}{c}{(8)} & \multicolumn{1}{c}{(9)} & \multicolumn{1}{c}{(10)} \\ 
\hline 

IC\,0480 & Sbc & 0.80 $\pm$ 0.01 & 8.49 $\pm$ 0.05 & 10.27 $\pm$ 0.13 & 9.55 $\pm$ 0.02 & 0.11 $\pm$ 0.10 & 3.08 $\pm$ 0.32 & 2.23 $\pm$ 0.43 & 2.58 $\pm$ 0.41 \tabularnewline 
 IC\,0944 & Sa & 0.75 $\pm$ 0.01 & 8.52 $\pm$ 0.06 & 11.26 $\pm$ 0.10 & 10.00 $\pm$ 0.02 & 0.41 $\pm$ 0.15 & 5.06 $\pm$ 0.15 & 5.16 $\pm$ 0.90 & 8.70 $\pm$ 0.79 \tabularnewline 
 IC\,2247 & Sbc & 0.72 $\pm$ 0.01 & 8.51 $\pm$ 0.04 & 10.44 $\pm$ 0.11 & 9.47 $\pm$ 0.02 & 0.23 $\pm$ 0.15 & 2.62 $\pm$ 0.13 & 2.91 $\pm$ 0.79 & 2.79 $\pm$ 0.46 \tabularnewline 
 IC\,2487 & Sb & 0.63 $\pm$ 0.01 & 8.52 $\pm$ 0.05 & 10.59 $\pm$ 0.12 & 9.34 $\pm$ 0.04 & 0.17 $\pm$ 0.08 & 3.83 $\pm$ 0.09 & 3.82 $\pm$ 1.03 & 5.36 $\pm$ 0.54 \tabularnewline 
 NGC\,2253 & Sc & 0.43 $\pm$ 0.01 & 8.59 $\pm$ 0.04 & 10.81 $\pm$ 0.11 & 9.62 $\pm$ 0.02 & 0.50 $\pm$ 0.06 & 2.48 $\pm$ 0.18 & 2.83 $\pm$ 0.85 & 1.82 $\pm$ 0.52 \tabularnewline 
 NGC\,2347 & Sbc & 0.63 $\pm$ 0.01 & 8.57 $\pm$ 0.04 & 11.04 $\pm$ 0.10 & 9.56 $\pm$ 0.02 & 0.54 $\pm$ 0.07 & 2.16 $\pm$ 0.06 & 2.45 $\pm$ 0.68 & 1.37 $\pm$ 0.35 \tabularnewline 
 NGC\,2410 & Sb & 0.89 $\pm$ 0.03 & 8.52 $\pm$ 0.05 & 11.03 $\pm$ 0.10 & 9.66 $\pm$ 0.03 & 0.55 $\pm$ 0.11 & 3.22 $\pm$ 0.13 & 4.09 $\pm$ 1.29 & 3.42 $\pm$ 0.19 \tabularnewline 
 NGC\,2730 & Sd & 0.79 $\pm$ 0.02 & 8.45 $\pm$ 0.04 & 10.13 $\pm$ 0.09 & 9.00 $\pm$ 0.06 & 0.23 $\pm$ 0.06 &    \multicolumn{1}{c}{(3.80)$^{a}$}   &    \multicolumn{1}{c}{-}   & 11.61 $\pm$ 4.11 \tabularnewline 
 NGC\,4644 & Sb & 1.30 $\pm$ 0.04 & 8.59 $\pm$ 0.04 & 10.68 $\pm$ 0.11 & 9.20 $\pm$ 0.05 & 0.09 $\pm$ 0.09 & 2.64 $\pm$ 0.18 & 7.18 $\pm$ 3.37 & 5.26 $\pm$ 0.80 \tabularnewline 
 NGC\,4711 & SBb & 0.93 $\pm$ 0.05 & 8.60 $\pm$ 0.04 & 10.58 $\pm$ 0.09 & 9.18 $\pm$ 0.05 & 0.08 $\pm$ 0.07 & 3.01 $\pm$ 0.11 & 5.59 $\pm$ 5.41 & 3.13 $\pm$ 0.68 \tabularnewline 
 NGC\,5056 & Sc & 1.09 $\pm$ 0.06 & 8.49 $\pm$ 0.03 & 10.85 $\pm$ 0.09 & 9.45 $\pm$ 0.04 & 0.57 $\pm$ 0.06 & 2.96 $\pm$ 0.08 & 4.37 $\pm$ 1.60 & 4.68 $\pm$ 0.59 \tabularnewline 
 NGC\,5614 & Sab & 1.00 $\pm$ 0.81 & 8.55 $\pm$ 0.06 & 11.22 $\pm$ 0.09 & 9.84 $\pm$ 0.01 & 0.20 $\pm$ 0.11 & 2.31 $\pm$ 0.21 & 1.04 $\pm$ 0.50 & 3.04 $\pm$ 1.04 \tabularnewline 
 NGC\,5908 & Sb & 1.01 $\pm$ 0.12 & 8.54 $\pm$ 0.05 & 10.95 $\pm$ 0.10 & 9.94 $\pm$ 0.01 & 0.36 $\pm$ 0.08 & 3.21 $\pm$ 0.07 & 3.25 $\pm$ 0.48 & 2.32 $\pm$ 0.24 \tabularnewline 
 NGC\,5980 & Sbc & 0.77 $\pm$ 0.01 & 8.58 $\pm$ 0.03 & 10.81 $\pm$ 0.10 & 9.70 $\pm$ 0.02 & 0.71 $\pm$ 0.06 & 2.37 $\pm$ 0.05 & 2.60 $\pm$ 0.60 & 1.87 $\pm$ 0.30 \tabularnewline 
 NGC\,6060 & SABc & 0.82 $\pm$ 0.03 & 8.50 $\pm$ 0.08 & 10.99 $\pm$ 0.09 & 9.68 $\pm$ 0.03 & 0.62 $\pm$ 0.14 & 3.90 $\pm$ 0.21 & 6.09 $\pm$ 1.77 & 5.31 $\pm$ 1.07 \tabularnewline 
 NGC\,6168 & Sd & 0.67 $\pm$ 0.01 & 8.40 $\pm$ 0.03 & 9.94 $\pm$ 0.11 & 8.65 $\pm$ 0.06 & $-0.07$ $\pm$ 0.06 & 2.42 $\pm$ 0.40 &   \multicolumn{1}{c}{-}   & 1.68 $\pm$ 0.53 \tabularnewline 
 NGC\,6186 & Sa & 0.88 $\pm$ 0.04 & 8.59 $\pm$ 0.04 & 10.62 $\pm$ 0.09 & 9.46 $\pm$ 0.02 & 0.30 $\pm$ 0.06 & 2.43 $\pm$ 0.11 & 2.25 $\pm$ 0.45 & 1.66 $\pm$ 0.40 \tabularnewline 
 NGC\,6314 & Sa & 0.54 $\pm$ 0.01 & 8.49 $\pm$ 0.06 & 11.21 $\pm$ 0.09 & 9.57 $\pm$ 0.03 & 0.00 $\pm$ 0.28 & 3.77 $\pm$ 0.21 & 2.25 $\pm$ 0.08 & 0.97 $\pm$ 0.18 \tabularnewline 
 NGC\,6478 & Sc & 0.62 $\pm$ 0.01 & 8.56 $\pm$ 0.04 & 11.27 $\pm$ 0.10 & 10.14 $\pm$ 0.02 & 1.00 $\pm$ 0.07 & 6.23 $\pm$ 0.27 & 6.60 $\pm$ 1.13 & 15.99 $\pm$ 4.00 \tabularnewline 
 NGC\,7738 & Sb & 0.70 $\pm$ 0.03 & 8.56 $\pm$ 0.06 & 11.21 $\pm$ 0.11 & 9.99 $\pm$ 0.01 & 1.18 $\pm$ 0.09 & 2.30 $\pm$ 0.24 & 1.68 $\pm$ 0.54 & 1.14 $\pm$ 0.20 \tabularnewline 
 UGC\,00809 & Sc & 0.68 $\pm$ 0.01 & 8.41 $\pm$ 0.03 & 10.00 $\pm$ 0.13 & 8.92 $\pm$ 0.07 & $-0.14$ $\pm$ 0.08 & 3.84 $\pm$ 0.16 & 6.14 $\pm$ 3.15 & 2.99 $\pm$ 0.36 \tabularnewline 
 UGC\,03253 & Sb & 1.21 $\pm$ 0.03 & 8.51 $\pm$ 0.07 & 10.63 $\pm$ 0.11 & 8.88 $\pm$ 0.06 & 0.23 $\pm$ 0.11 & 2.42 $\pm$ 0.09 & 5.14 $\pm$ 1.58 & 3.16 $\pm$ 1.03 \tabularnewline 
 UGC\,03539 & Sbc & 1.25 $\pm$ 0.07 & 8.39 $\pm$ 0.07 & 9.84 $\pm$ 0.13 & 9.11 $\pm$ 0.03 & $-0.17$ $\pm$ 0.09 & 1.46 $\pm$ 0.02 & 1.58 $\pm$ 1.03 & 1.62 $\pm$ 0.15 \tabularnewline 
 UGC\,04029 & Sbc & 0.78 $\pm$ 0.02 & 8.48 $\pm$ 0.08 & 10.38 $\pm$ 0.10 & 9.37 $\pm$ 0.03 & 0.18 $\pm$ 0.09 & 3.38 $\pm$ 0.16 & 4.03 $\pm$ 0.97 & 4.33 $\pm$ 0.34 \tabularnewline 
 UGC\,04132 & Sbc & 0.99 $\pm$ 0.33 & 8.54 $\pm$ 0.04 & 10.94 $\pm$ 0.12 & 10.02 $\pm$ 0.01 & 0.96 $\pm$ 0.07 & 3.63 $\pm$ 0.16 & 3.13 $\pm$ 0.62 & 4.42 $\pm$ 0.49 \tabularnewline 
 UGC\,05108 & SBab & 1.16 $\pm$ 0.03 & 8.50 $\pm$ 0.06 & 11.11 $\pm$ 0.11 & 9.75 $\pm$ 0.04 & 0.66 $\pm$ 0.12 & 3.79 $\pm$ 0.10 & 2.75 $\pm$ 0.80 & 2.72 $\pm$ 0.28 \tabularnewline 
 UGC\,05598 & Sbc & 0.54 $\pm$ 0.01 & 8.45 $\pm$ 0.05 & 10.40 $\pm$ 0.12 & 9.17 $\pm$ 0.06 & 0.15 $\pm$ 0.09 & 3.09 $\pm$ 0.21 & 2.68 $\pm$ 0.72 & 4.59 $\pm$ 0.51 \tabularnewline 
 UGC\,09542 & Sc & 0.46 $\pm$ 0.01 & 8.49 $\pm$ 0.05 & 10.53 $\pm$ 0.13 & 9.31 $\pm$ 0.05 & 0.27 $\pm$ 0.09 & 3.45 $\pm$ 0.10 & 5.44 $\pm$ 2.24 & 5.96 $\pm$ 1.05 \tabularnewline 
 UGC\,09873 & Sc & 0.76 $\pm$ 0.02 & 8.46 $\pm$ 0.05 & 10.21 $\pm$ 0.10 & 9.08 $\pm$ 0.07 & 0.10 $\pm$ 0.09 & 3.69 $\pm$ 0.14 & 2.86 $\pm$ 0.94 & 2.97 $\pm$ 0.27 \tabularnewline 
 UGC\,09892 & Sb & 1.04 $\pm$ 0.30 & 8.48 $\pm$ 0.05 & 10.48 $\pm$ 0.10 & 9.17 $\pm$ 0.05 & $-0.03$ $\pm$ 0.08 & 2.90 $\pm$ 0.12 & 5.72 $\pm$ 2.05 & 4.78 $\pm$ 0.61 \tabularnewline 
 UGC\,10123 & Sab & 0.72 $\pm$ 0.01 & 8.54 $\pm$ 0.03 & 10.30 $\pm$ 0.10 & 9.48 $\pm$ 0.02 & 0.21 $\pm$ 0.07 & 1.62 $\pm$ 0.11 & 2.23 $\pm$ 0.59 & 2.19 $\pm$ 0.20 \tabularnewline 
 UGC\,10205 & Sa & 0.97 $\pm$ 0.07 & 8.49 $\pm$ 0.04 & 11.08 $\pm$ 0.10 & 9.60 $\pm$ 0.04 & 0.38 $\pm$ 0.20 & 3.12 $\pm$ 0.09 & 2.94 $\pm$ 0.84 & 2.01 $\pm$ 0.06 \tabularnewline 
 UGC\,10384 & Sab & 0.70 $\pm$ 0.01 & 8.50 $\pm$ 0.05 & 10.33 $\pm$ 0.14 & 9.10 $\pm$ 0.02 & 0.65 $\pm$ 0.06 & 1.53 $\pm$ 0.10 & 1.77 $\pm$ 0.29 & 1.84 $\pm$ 0.16 \tabularnewline 
 UGC\,10710 & Sb & 0.66 $\pm$ 0.01 & 8.52 $\pm$ 0.05 & 10.92 $\pm$ 0.09 & 9.88 $\pm$ 0.04 & 0.50 $\pm$ 0.10 & 5.15 $\pm$ 0.42 & 4.39 $\pm$ 0.96 & 4.62 $\pm$ 0.55 \tabularnewline

 \hline 
 {\bf{Notes.}} \tabularnewline
 \multicolumn{10}{l}{Column 1: galaxy name; Column 2: Hubble type; Column 3: ratio of vertical to radial velocity dispersion calculated from $\beta_{z}$ derived by \citetalias{kal17b}; Column 4: } \\
 \multicolumn{10}{l}{metallicity; Column 5: stellar mass; Column 6: molecular gas mass; Column 7: star formation rate; Column 8: stellar scale length; Column 9: molecular } \\
 \multicolumn{10}{l}{gas scale length; Column 10: star formation scale length.  Column 2 and 4-10 are from \citetalias{bol17}. }\\
 \multicolumn{10}{l}{$^{a}$the scale length for NGC 2730 is estimated using $l_{*} = R_{\rm e} / 1.68$. }

\end{tabular}
\end{table*}

%Tale 0

\section{METHOD}

We derive the radial velocity dispersion \sigmaR maps from \sigmao maps using the thin-disc approach \citep[see, e.g.,][]{bin98}. Firstly, the line-of-sight velocity dispersion is expressed in terms of the radial \sigmaR, tangential \sigmap and vertical \sigmaz dispersion components by the general formula:
\begin{equation}
 \sigma_{\rm{los}}^2 = \left( \sigma_{R}^2 \sin^2 \phi + \sigma_{\phi}^2 \cos^2 \phi \right) \sin^2 i + \sigma_{z}^2 \cos^2 i,
\end{equation}
which requires the inclination angle of the galaxy $i$ and the position angle of the galaxy $\phi$  \citep[e.g.,][]{bin98}.  \citet[][]{rom16} define two parameters (based on the axis ratios of the dispersion anisotropy components): $A=$\,\sigmap/\sigmaR\,and $B =$\,\sigmaz/\sigmaR\,in order to rewrite the above equation in the form:
\begin{equation}
 \sigma_{R} = \sigma_{\rm{los}} \left[ \left( \sin^2 \phi + A^2 \cos^2 \phi \right) \sin^2 i + B^2 \cos^2 i \right]^{-1/2}. 
\end{equation}
\\

Following the epicyclic approximation of an axisymmetric disc with approximately circular orbits $A \approx \kappa / 2\Omega$ \citep[e.g.,][]{bin08}, where $\Omega$ is the angular frequency and $\kappa$ the epicyclic frequency.  Each of these parameters can be determined from circular velocity $v_c(R)$ as follows: $\Omega = v_c(R)/R$ and $\kappa = \sqrt{ R\,{\rm{d}}\Omega^2 / {\rm{d}}R + 4\Omega^2 }$.

\begin{figure}
 %\centering
 \includegraphics[width=0.45\textwidth]{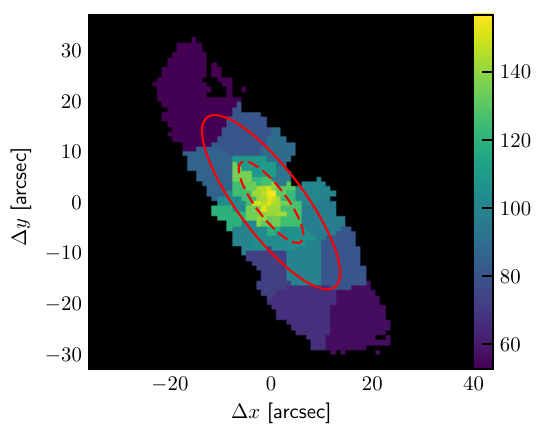}
 \caption{ A map of the stellar radial velocity dispersion \sigmaR in NGC 2410, the colourbar represents \sigmaR values in units of \kph.  The solid circle represents one effective radius $R_{\rm e}$ and the dashed line represents one stellar scale length $l_{\rm s}$.
 }
 \label{fig:dispmap}
\end{figure}

We use \citetalias{kal17b} circular-speed curves to calculate $\kappa$ and $\Omega$, from which we calculate $A$, and use their dispersion anisotropy parameter $\beta_{z} = 1 - \sigma_{z}^2 / \sigma_{R}^2$.  We calculate $B$ using $B = \sqrt{1-\beta_{z}^2}$\,.  Therefore we have the necessary parameters to calculate \sigmaR from \sigmao using Equation 1 thus we use maps of $A$, $B$, $\phi$ and \sigmao to calculate \sigmaR and produce maps of it for each galaxy.  An example of a \sigmaR map is showed in Figure \ref{fig:dispmap} .

\begin{figure*}
 \centering
 \includegraphics[width=0.95\textwidth]{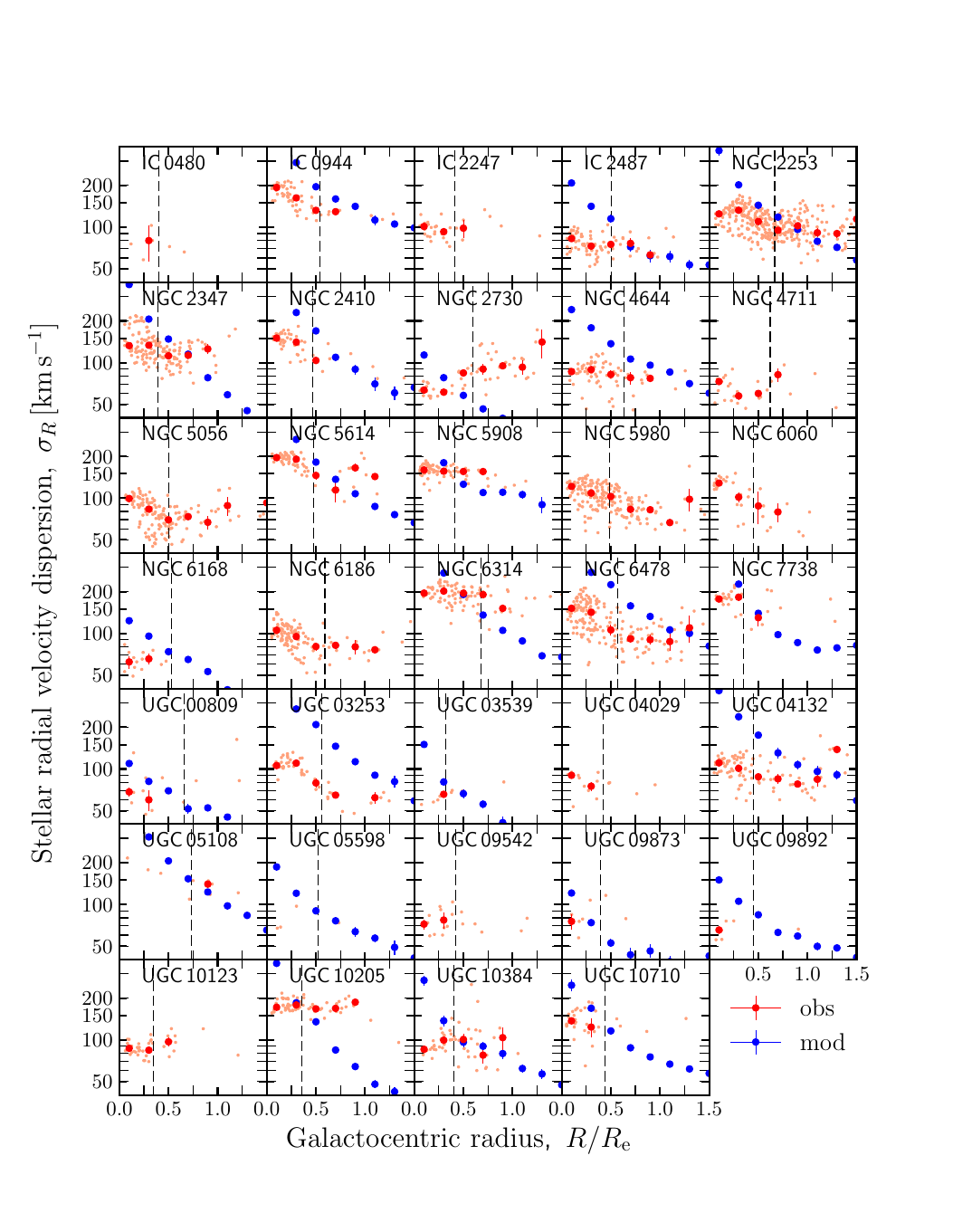}
 \caption{Stellar radial velocity dispersion \sigmaR as a function of galactocentric radius.  Light red data points show individual \sigmaR measurements based on line-of-sight velocity dispersion measurements, the dark red data points are the median and associated error for \sigmaR data in $0.2$\,$R_e$ bins. Blue points are the median and associated uncertainty of model-based velocity dispersions in $0.2$\,$R_e$ bins.  The vertical dashed lines indicate the stellar scale length.
 }
 \label{fig:disps}
\end{figure*}

\begin{figure*}
 \centering
 \includegraphics[width=\textwidth]{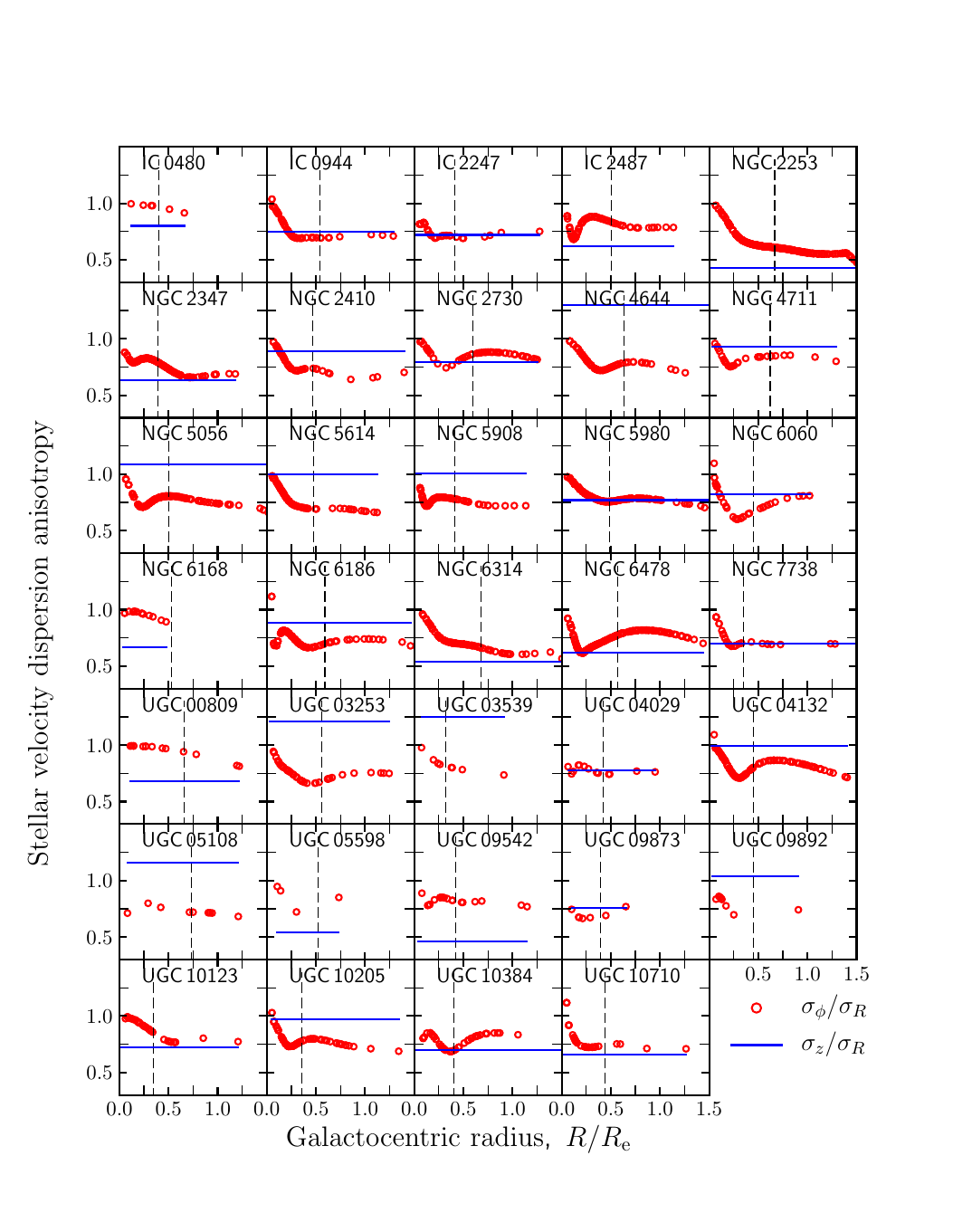}
 \caption{Stellar velocity dispersion anisotropy parameters as a function of galactocentric radius.  The red circles are $A=\sigma_{\phi}/\sigma_{R}$ calculated at the galactocentric radius of each \sigmao data point, the blue lines show the constant $B=\sigma_{z}/\sigma_{R}$ calculated from $\beta_{z}$ values derived by \citetalias{kal17b}.  The vertical dashed lines indicate the stellar scale length.
 }
 \label{fig:dispasym}
\end{figure*}

We use \sigmao maps to mask out unreliable \sigmaR values by imposing $40$\,\kph as a lower limit on our \sigmao, because \mbox{\citetalias{fal17}} compared their \sigmao values with higher resolution \sigmao observations and found that the CALIFA \sigmao values and their associated uncertainties are highly unreliable for $\sigma < 40$\,\kph.  We further apply a cut-off to exclude data with relative uncertainties greater than $20$\%.  This value is based on the median relative uncertainty of data with $\sigma \sim 40$\,\kph being $20$\% (\citetalias{fal17}).  After we apply these we are left with reliable \sigmaR maps for each galaxy. 

We derive the radial profiles of \sigmaR by dividing \sigmaR maps into tilted rings that are circular in the plane of the galaxy. Each tilted ring is defined by a kinematically derived (where possible) inclination and position angle taken from \mbox{\citetalias{bol17}}, and the galaxy center is defined as the photometric center adopted by \mbox{\citetalias{fal17}} in their \sigmao maps \mbox{\citep{hus13}}.  Figure \mbox{\ref{fig:dispmap}} shows an example of azimuthal rings defined by the effective radius and stellar scale length.  Then we calculate the median and its associated uncertainty for each radial bin of width $0.2\,R_{\rm{e}}$. Only annuli that contain more than 2 data points are used for the \sigmaRR calculations.  In such data some individual rings contain few data points and some have a large fraction of outliers, therefore we use the median and its associated uncertainty for robust statistical measures \citep*[e.g.,][]{rou91,mul00,rom04,hub09,fei12}.  In our study we calculate the uncertainty of the median by using the median absolute deviation (MAD):
\begin{equation}
\Delta X_{\rm{med}} = 1.858 \times \rm{MAD}/\sqrt{N}, 
\end{equation}
\begin{equation}
 \rm{MAD} = \rm{median}\{ | X_{i} - X_{\rm{med}} | \} ,
\end{equation}
where $X_{i}$ are individual measurements, $X_{\rm{med}}$ is their median value and $N$ is the number of pixels (Voronoi bin centers) in each ring where there are detections.  These equations are robust counterparts of the mean uncertainty formula which uses the standard deviation ($SD$): $\Delta X_{\rm{mean}} = SD/\sqrt{N}$ \citep[][]{mul00}.  We use these medians and associated uncertainties to determine the final radial profiles for \sigmaR and $A$.  The uncertainties do not take into account the covariance between bins.

The third step of the data analysis is to compare \sigmaR with modeled radial dispersions \sigmam.  We use the common approach used by \citep[][hereafter \citetalias{ler08}]{ler08} to determine \sigmam (see Appendix B.3 of \citetalias{ler08}) :
\begin{equation} {\label{eq:smod}}
\sigma_{\rm{mod}} = \frac{1}{0.6} \sqrt \frac{2\pi G l_{\star}}{7.3} \Sigma_{\star}^{0.5},
\end{equation}
where $l_{\star}$ is the stellar exponential scale length and $\Sigma_{\star}$ is the stellar surface density.  

This model assumes that the exponential scale height of a galaxy does not vary with radius, the flattening ratio between the scale height and scale length is $7.3$ \citep[][]{kre02}, that discs are in hydrostatic equilibrium and that they are isothermal in the z-direction \citep[e.g.,][]{vdkru81,vdkru88} and that \sigmaz/\sigmaR\,$=0.6$ \citep[e.g.,][]{sha03}.  We investigate the effects of the flattening ratio and \sigmaz/\sigmaR\, assumptions on our analysis in Section 5.2\,.  For each galaxy in our subsample we take the \Sigmas map and $l_{\star}$ values (from \citetalias{bol17}) and use Equation 5 to derive a map of \sigmam.  Then we divide the \sigmam map into tilted rings that are circular in the plane of the galaxy.  And we determine the radial profile by calculating the median and its associated uncertainty for each radial bin of width $0.2\,R_{\rm{e}}$.  The outputs of this procedure are maps and radial profiles of \sigmam for each galaxy in our subsample.

\section{RESULTS}

\subsection{Radial profiles}

In Figure \ref{fig:disps} we show the \sigmaR of each Voronoi bin and as a function of galactocentric radius (\sigmaRR) for each galaxy in our sample.  There are large variations in the radial behaviour of \sigmaR between galaxies, but the general trend is of decreasing \sigmaR with increasing $R$.   

Comparisons between \sigmaR\, and \sigmam\, are displayed in Figure \ref{fig:disps}.  The radial behaviour of \sigmam is dominated by the typically exponential smooth decrease of \Sigmas and in the figure we see a far more pronounced decrease of \sigmam with increasing $R$ than for \sigmaR.  Figure \ref{fig:disps} shows that \sigmam overestimates \sigmaR at low $R$, and in general at $R = l_{\star}$ we find that \sigmaR $<$ \sigmam\,.  The data and shallower decline result in a switch-over at larger $R$ where \sigmaR $\geq$ \sigmam\,.  However, due to the sparseness of \sigmaR data at large $R$ we cannot conclude that this is the general behaviour.

Figure \ref{fig:dispasym} shows the radial behaviour of $A$ and $B$ parameters calculated from kinematic parameters derived by \citetalias{kal17b}.  Parameter $B$ is constant due to the assumption of a constant $\beta_{z}$ by \citetalias{kal17b} and $A$ typically decreases with increasing $R$ from a maximum $\sim 1$. There is a large variation in $B$ between galaxies, ranging between $0.4$ and $1.3$, which is larger than found in previous studies and typically used in models \citep[][]{sha03,ler08,rom15,rom16,pin18}.

\begin{figure}
 \centering
 \includegraphics[width=0.45\textwidth]{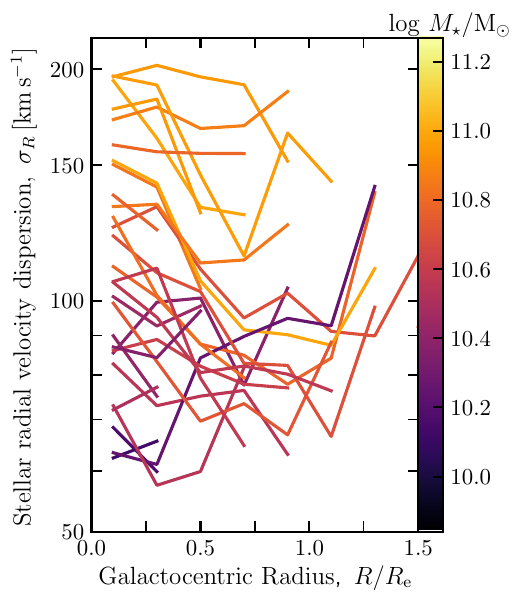}
 \caption{The stellar radial velocity dispersion as a function of galactocentric radius \sigmaRR for each galaxy. Galaxies are colour coded according to $M_{\star}$. The \sigmaRR values plotted are the medians of \sigmaR in $0.2$\,$R_e$ bins.
 }
 \label{fig:disprad}
\end{figure}

We now study the relationship between \sigmaRR and galaxy properties.  The data does not extend far out enough to determine whether the radial behaviour of \sigmaR correlates with any of the properties.  In Figure \mbox{\ref{fig:disprad}} we plot \sigmaR as a function of galactocentric radius and $M_{\star}$.  It should be noted that measurements of $M_{\star}$ are limited to within the $74''\times 64''$ field of view of the CALIFA observations, and \mbox{\citep{gon14}} showed that on average this can underestimate the total $M_{\star}$ by $8$\%. The are large \sigmaR variations between and within galaxies as in Figure \ref{fig:disps}.  However, from the figure we see that galaxies with higher $M_{\star}$ tend to have larger \sigmaR\,.  When we compare the radial behaviour of \sigmaR\, and other properties we see not correlations, however, the relationships between different parameters and \sigmaR\, are discussed in more detail in the following section.

\begin{figure*}
 \centering
 \includegraphics[width=\textwidth]{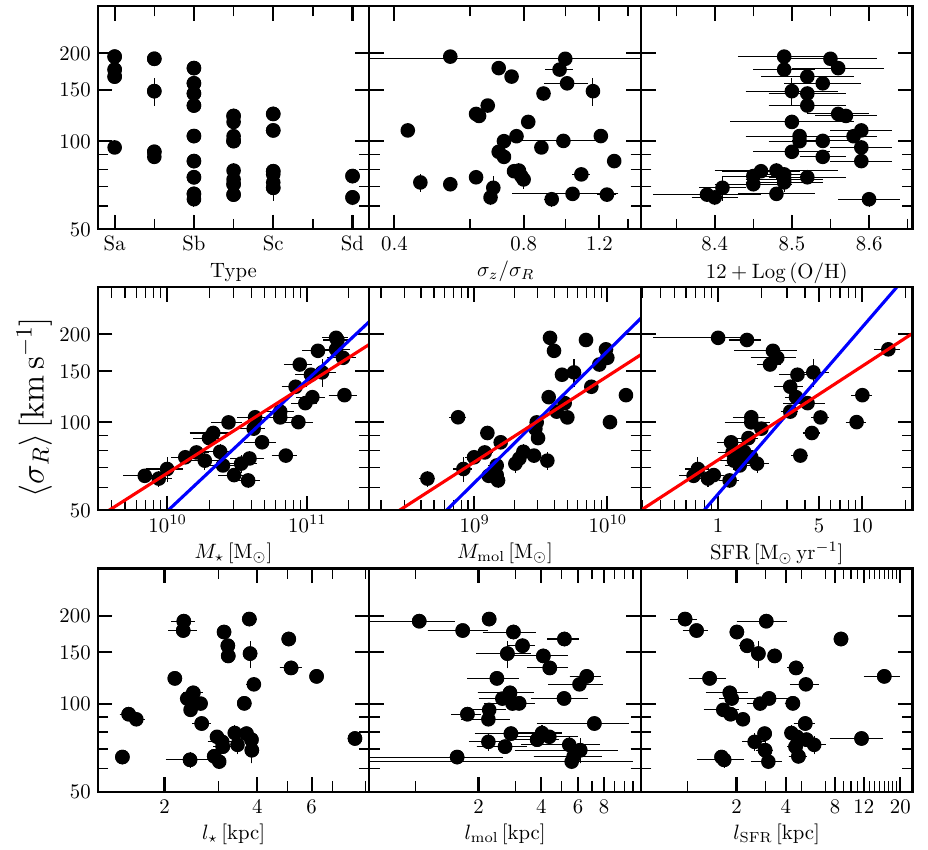}
 \caption{The radial average of the stellar radial velocity dispersion \sigmaRR over $1\,R_{\rm{e}}$, robustly estimated via the median, plotted as a function of Hubble type, $B$, metallicity, $M_{\star}$, $M_{\rm{mol}}$, SFR, $l_{\star}$, $l_{\rm{mol}}$ and $l_{\rm{SFR}}$.  The red lines represent the best-fitting lines using a robust median-based fit method and the blue lines represent the best-fitting lines from ODR least-squares fitting.
 }
 \label{fig:meddispprop}
\end{figure*}

\subsection{Correlations}

We want to quantify the relationships between \sigmaR\, and different parameters over a physically significant region of the galaxy and hence calculate the radial average of \sigmaRR over one effective (half-light) radius, robustly estimated via the median $ \langle \sigma_{R} \rangle $ and its associated uncertainty for each galaxy.  These are derived using the same method as \sigmaRR but with a ring width equal to $1\,R_{\rm{e}}$.  We do not apply any corrections for galaxies whose data does not extend to $1\,R_{\rm{e}}$.  The $\langle \sigma_R \rangle$ for each galaxy are plotted against various properties in Figure \ref{fig:meddispprop}.  The Pearson's and Spearman's correlation coefficients ($r_{\rm P}$ and $r_{\rm S}$ respectively), their corresponding $\rm p$-values (which indicate the probability of a null-hypothesis) and best-fitting linear parameters of each \medsigmaR -- parameter plot are shown in Table \ref{tab:corr}.  Linear fits were were parametrized as follows:  $\log \langle \sigma_{R} \rangle = a \log X + b$ for fits performed using the robust median method and $\log \langle \sigma_{R} \rangle = c \log X + d$ for fits performed using the least-squares orthogonal distance regression (ODR) method \mbox{\citep[see, e.g.,][]{pre92}}.  The latter method takes into account uncertainties of both variables whereas the former does not take into account any uncertainties but is a more robust fitting method.  The best-fitting lines and parameters are only shown in Figure \ref{fig:meddispprop} and Table \ref{tab:corr} for cases where there is a strong and significant correlation between variables, we define this case as $|r|> 0.5$ and ${\rm p} < 0.05$.  The relative strengths and significances of correlations are consistent whether Pearson's or Spearman's correlation coefficients are used.

\begin{table*}
\caption{Correlation coefficients and best-fitting parameters for \sigmaR versus galaxy properties.}
\label{tab:corr}
\begin{tabular}{lcccccccccc}
\hline 
\multicolumn{1}{l}{Property} & \multicolumn{1}{c}{$r_{\rm{P}}$} & \multicolumn{1}{c}{$p_{\rm{P}}$} & \multicolumn{1}{c}{$r_{\rm{S}}$} & \multicolumn{1}{c}{$p_{\rm{S}}$} & \multicolumn{1}{c}{$a$} & \multicolumn{1}{c}{$b$}  & \multicolumn{1}{c}{$c$} & \multicolumn{1}{c}{$d$} & \multicolumn{1}{c}{$\Delta$}  \\
\multicolumn{1}{l}{} & \multicolumn{1}{c}{} & \multicolumn{1}{c}{}  & \multicolumn{1}{c}{} & \multicolumn{1}{c}{} & \multicolumn{1}{c}{} & \multicolumn{1}{c}{} & \multicolumn{1}{c}{} & \multicolumn{1}{c}{} & \multicolumn{1}{c}{}  \\
%\hline
\multicolumn{1}{l}{(1)} & \multicolumn{1}{c}{(2)} & \multicolumn{1}{c}{(3)} & \multicolumn{1}{c}{(4)} & \multicolumn{1}{c}{(5)} & \multicolumn{1}{c}{(6)} & \multicolumn{1}{c}{(7)} & \multicolumn{1}{c}{(8)} & \multicolumn{1}{c}{(9)} & \multicolumn{1}{c}{(10)} \\ 
\hline 

Hubble Stage (T) & \multicolumn{1}{c}{$-0.58$\,\,\,\,} & $3.5\times10^{-4}$ & \multicolumn{1}{l}{$-0.51$\,\,\,\,} & $1.8\times10^{-3}$ & - & - & - & - & - \tabularnewline 
$\sigma_{z}/\sigma_{R}$ & \multicolumn{1}{c}{0.00} & {$1.0$\,\,} & 0.00 & $9.9\times10^{-1}$ & - & - & - & - & - \tabularnewline 
12+Log(O/H) & 0.32 & $6.2\times10^{-2}$ & 0.44 & $1.0\times10^{-2}$ & - & - & - & - & - \tabularnewline 
$M_{\star}$ [M$_{\odot}$] & 0.82 & \multicolumn{1}{l}{$2.2\times10^{-9}$} & 0.86 & \multicolumn{1}{l}{\, $1.0\times10^{-10}$} & 0.30 & \multicolumn{1}{l}{\,$-1.22$\,\,\,\,} & 0.45 $\pm$ 0.05 & \multicolumn{1}{l}{$-2.78$ $\pm$ 0.51\,\,\,\,} & 0.10 \tabularnewline 
$M_{\rm{mol}}$ [M$_{\odot}$] & 0.69 & $5.6\times10^{-6}$ & 0.77 & $1.0\times10^{-7}$ & 0.29 & \multicolumn{1}{l}{\,$-0.78$\,\,} & 0.45 $\pm$ 0.06 & \multicolumn{1}{l}{$-2.26$ $\pm$ 0.62\,\,} & 0.12 \tabularnewline 
SFR [\msolyr] & 0.42 & $1.3\times10^{-2}$ & 0.60 & $1.8\times10^{-4}$ & 0.32 & 1.87 & 0.57 $\pm$ 0.11 & \multicolumn{1}{l}{\,\,\,\,\,1.76 $\pm$ 0.05} & 0.18 \tabularnewline 
$l_{\star}$ [kpc] & 0.07 & $7.0\times10^{-1}$ & 0.10 & $5.6\times10^{-1}$ & - & - & - & - & - \tabularnewline 
$l_{\rm{mol}}$ [kpc] & \multicolumn{1}{l}{$-0.28$\,\,\,\,} & $1.1\times10^{-1}$ & \multicolumn{1}{l}{$-0.20$\,\,\,\,} & $2.7\times10^{-1}$ & - & - & - & - & - \tabularnewline 
$l_{\rm{SFR}}$ [kpc] & $-0.09$\,\,\,\, & $6.0\times10^{-1}$ & \multicolumn{1}{l}{$-0.19$\,\,\,\,} & $2.9\times10^{-1}$ & - & - & - & - & - \tabularnewline 
 \hline 
 {\bf{Notes.}} \tabularnewline
 \multicolumn{10}{l}{Column 1: galaxy property ; Column 2: Pearson's rank correlation coefficient; Column 3: $p$-value for Pearson's rank correlation; Col-} \\
 \multicolumn{10}{l}{umn 4: Spearman's rank correlation coefficient; Column 5: $p$-value for Spearman's  rank correlation; Column 6,7: $a$ and $b$ parameters } \\
 \multicolumn{10}{l}{ from the robust median-based fit $\log \langle \sigma_{R} \rangle$ $= a\log X + b$\,, where X denotes galaxy property; Column 8,9: $c$ and $d$ parameters from the} \\
 \multicolumn{10}{l}{ ODR fit $\log \langle \sigma_{R} \rangle = c \log X + d$; Column 10: rms scatter of scaling relations.   } \\
 \multicolumn{10}{l}{}
\end{tabular}
\end{table*}

In Figure \ref{fig:meddispprop} we see that \medsigmaR\, is correlated with $M_{\star}$, $M_{\rm mol}$ and SFR respectively.  This is confirmed by the correlation coefficients shown in Table \ref{tab:modelcomp} which range from $0.42$ (SFR) to $0.86$ ($M_{\star}$).  Among the galaxy properties, $M_{\star}$ has the strongest and most significant correlation with $\langle\,\sigma_{R}\,\rangle$, the correlation between them has $r_{\rm S}=0.86$ and ${\rm p_S}=1.0 \times 10^{-10}$.  The best-fitting linear relationship is $\log \langle\,\sigma_{R}\,\rangle = (0.45 \pm 0.05) \log M_{\star} + (-2.78 \pm 0.51)$ with a root mean squared (rms) scatter of $0.10$ dex ($26$\,\%); therefore $\langle\,\sigma_{R}\,\rangle \propto M_{\star}^{0.45}$.  $M_{\rm mol}$ has the next strongest and significant correlation ($r_{\rm S}=0.77 $ and ${\rm p_S}= 1.0 \times 10^{-7}$) followed by SFR ($r_{\rm S}=0.60 $ and ${\rm p_S}= 1.8 \times 10^{-4}$).  And their best-fitting relations have rms scatter values of $0.12$ and $0.18$ dex respectively.  The power law indices of the $M_{\star}$, $M_{\rm mol}$ and SFR relations are close to $0.5$ when uncertainties are taken into account, when no uncertainties are taken into account the indices are lower and range between $0.29$ and $0.32$.

We also see weak \medsigmaR\, correlations with Hubble type ($r_{\rm S}=-0.51 $) and metallicity ($r_{\rm S}=0.44 $), both have lower significance than the aforementioned properties, their p-values less than $0.05$\,.  The other parameters (\sigmaz/\sigmaR\,, $l_{\star}$, $l_{\rm{mol}}$, $l_{\rm{SFR}}$) are not correlated with \medsigmaR\,, their p-values are larger than 0.05\,.

 \begin{figure}
 \centering
 \includegraphics[width=0.45\textwidth]{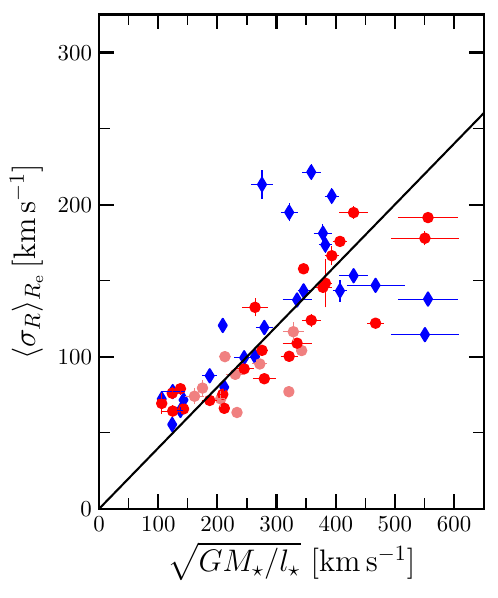}
 \caption{The radial averages of the stellar velocity dispersion \sigmaRR (red circles) and model-based velocity dispersion \sigmam$(R)$ (blue diamonds) over an effective radius $R_{\rm{e}}$, robustly estimated via the median, all plotted against the velocity scale: \gml\,.  Dark red points show the subsample of galaxies for which we calculated model-based velocity dispersions.  A \medsigmaR $= 0.4$\gml\, relation is shown by the black line.  
 }
 \label{fig:sigmamod}
\end{figure}

 \begin{figure}
 \centering
 \includegraphics[width=0.45\textwidth]{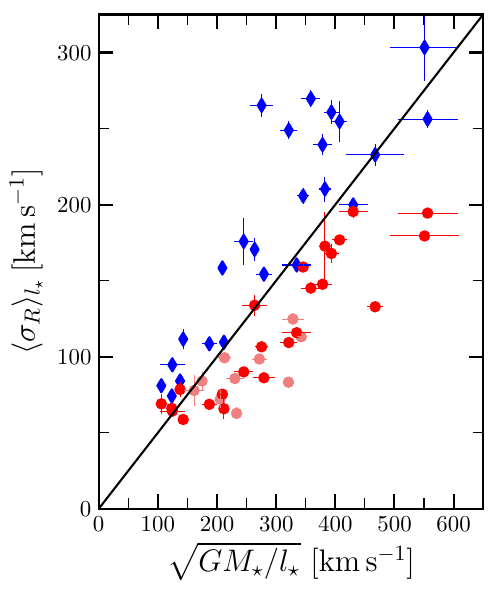}
 \caption{The radial averages of the stellar velocity dispersion \sigmaRR (red circles) and model-based velocity dispersion \sigmam$(R)$ (blue diamonds) over the stellar scale length $l_{\star}$, robustly estimated via the median, all plotted against the velocity scale: \gml\,. Dark red points show the subsample of galaxies for which we calculated model-based velocity dispersions.  A \medsigmaR $= 0.5$\gml\, relation is shown by the black line. 
 }
 \label{fig:sigmamodls}
\end{figure}

Finally we test the \sigmam model by determining the radial average of \sigmam$(R)$ over $1\,R_{\rm e}$, robustly estimated via the median and comparing it with the observation-based $\langle\,\sigma_{R}\,\rangle$ in Figure \ref{fig:sigmamod}.  We plot them against the velocity scale: $\sqrt {GM_{\star}/l_{\star}} $ determined from the global properties: $M_{\star}$ and $l_{\star}$.  This was done for the 24/34 galaxies in our sample which have \Sigmas maps available.  Figure \ref{fig:sigmamod} is consistent with the findings in Figure \ref{fig:disps} where we find that \sigmam $>$ \sigmaR in the inner regions, and the difference between them tends to decrease as $R$ increases.  The data used in Figure \ref{fig:sigmamod} are shown in Table \ref{tab:modelcomp}.  We see in Figure \ref{fig:sigmamod} and Table \ref{tab:modelcomp} that \medsigmam $>$\medsigmaR\, for most galaxies.  Figure \ref{fig:sigmamod} has a separatrix line of \medsigmaR $= 0.4$\gml, derived by taking the radial average of Equation {\ref{eq:smod}} over \mbox{$R_{\rm{e}}$}, which is where we expect the \citetalias{ler08} \sigmam values to lie.  \medsigmaR\,values lie on or below this line and \medsigmam\, data tend to lie on or above this relation.  Therefore \medsigmam\, does not accurately model \medsigmaR.  

The expected relation between \sigmaR\, and \gml\, requires that 1) \sigmaR\, follow an exponential decline with radius and 2) that the spatial bin size of data points be equal.  However, Figure \ref{fig:disps} shows that \sigmaRR has a wide range of shapes even though it tends to decline with radius.  Therefore it is not always declining exponentially and due to the nature of our data the second condition of equal spatial bin sizes is not satisfied either.  These are the likely reasons for \medsigmam\, not following a slope of $0.4$\,.  

The fact that \medsigmam\, overestimates \medsigmaR\, significantly in the inner stellar disc becomes even clearer if we consider  the radial average of \sigmaRR over one exponential scale length $l_{\star}$ ($R_{\rm e} \sim 1.68 l_{\star}$) in Figure \ref{fig:sigmamodls}.  We see that the data are further away from the expected relation.  The plot shows that \medsigmam\, overestimates the observationally based \medsigmaR\, within $l_{\star}$, the differences are larger than in Figure \ref{fig:sigmamod} and are greater than $50$\,\kph\, in the most extreme cases.  This comparison confirms that the difference between \medsigmaR\, and \medsigmam\, is largest at small radii.

%Table 3

\begin{table}
\caption{Observed versus model-based \medsigmaR\,.}
\label{tab:modelcomp}
\begin{tabular}{|l|c|c|r|}

\hline 
\multicolumn{1}{l}{Name} & \multicolumn{1}{c}{\medsigmaRobs} & \multicolumn{1}{c}{\medsigmaRmod} & \multicolumn{1}{c}{$\sqrt{GM_{\star}/l_{\star}}$}  \\

\multicolumn{1}{l}{} & \multicolumn{1}{c}{[\kph]}  & \multicolumn{1}{c}{[\kph]}  & \multicolumn{1}{c}{[\kph]}  \\

%\hline
\multicolumn{1}{l}{(1)} & \multicolumn{1}{c}{(2)} & \multicolumn{1}{c}{(3)} & \multicolumn{1}{c}{(4)}  \\ 
\hline 

IC\,0480 & \,\,\,74.0 $\pm$ 5.2 &    \multicolumn{1}{c}{\,\,\,\,\,\,-}   & 161.3 $\pm$ 16.9 \tabularnewline 
IC\,0944 & 166.4 $\pm$  6.4 & 205.6 $\pm$  5.4 & 393.4 $\pm$ 12.2 \tabularnewline 
IC\,2247 & 100.1 $\pm$  2.6 &    \multicolumn{1}{c}{\,\,\,\,\,\,-}   & 212.7 $\pm$ 10.8 \tabularnewline 
IC\,2487 & \,\,\,75.3 $\pm$  1.8 & 120.6 $\pm$  4.6 & 209.1 $\pm$ 5.5\,\,\,  \tabularnewline 
NGC\,2253 & 108.9 $\pm$  1.9 & 137.3 $\pm$  2.9 & 334.8 $\pm$ 24.5 \tabularnewline 
NGC\,2347 & 122.0 $\pm$  2.5 & 146.9 $\pm$  4.0 & 467.4 $\pm$ 13.7 \tabularnewline 
NGC\,2410 & 146.6 $\pm$  3.2 & 181.1 $\pm$  5.9 & 378.5 $\pm$ 15.7 \tabularnewline 
NGC\,2730 &   \,\,\,76.0 $\pm$  4.0  &   \,\,\,55.4 $\pm$  1.2  &  (123.6)$^a$\,\,\,\,\,\,  \tabularnewline 
NGC\,4644 & \,\,\,85.5 $\pm$  1.9 &  119.2 $\pm$  2.8 & 279.4 $\pm$ 19.3 \tabularnewline 
NGC\,4711 & \,\,\,63.3 $\pm$  3.7 &   \multicolumn{1}{c}{\,\,\,\,\,\,-}  & 233.2 $\pm$ 8.8\,\,\, \tabularnewline 
NGC\,5056 & \,\,\,77.0 $\pm$  2.0 &   \multicolumn{1}{c}{\,\,\,\,\,\,-}  & 320.9 $\pm$ 9.1\,\,\, \tabularnewline 
NGC\,5614 & 191.5 $\pm$  3.2 & 137.9 $\pm$  1.8 & 556.1 $\pm$ 50.8 \tabularnewline 
NGC\,5908 & 157.8 $\pm$  1.7 &   143.3 $\pm$  2.3  & 345.7 $\pm$ 8.2\,\,\, \tabularnewline 
NGC\,5980 & 104.0 $\pm$  2.3 &   \multicolumn{1}{c}{\,\,\,\,\,\,-}  & 342.4 $\pm$ 7.9\,\,\, \tabularnewline 
NGC\,6060 & 116.4 $\pm$  6.1 &   \multicolumn{1}{c}{\,\,\,\,\,\,-}  & 328.4 $\pm$ 17.9 \tabularnewline 
NGC\,6168 & \,\,\,64.3 $\pm$  4.0 & \,\,\,77.1 $\pm$  1.8 & 124.5 $\pm$ 20.6 \tabularnewline 
NGC\,6186 & \,\,\,95.2 $\pm$  2.0 &   \multicolumn{1}{c}{\,\,\,\,\,\,-}  & 271.7 $\pm$ 12.5 \tabularnewline 
NGC\,6314 & 194.8 $\pm$  4.2 & 153.3 $\pm$  4.1 & 430.3 $\pm$ 24.2 \tabularnewline 
NGC\,6478 & 124.0 $\pm$  4.2 & 221.4 $\pm$  5.3 & 358.7 $\pm$ 15.9 \tabularnewline 
NGC\,7738 & 177.9 $\pm$  4.6 & \,114.5 $\pm$ 3.0 & 550.9 $\pm$ 57.7 \tabularnewline 
UGC\,00809 & \,\,\,69.3 $\pm$  6.8 & \,\,\,72.2 $\pm$  2.1 & 105.9 $\pm$ 4.6\,\,\, \tabularnewline 
UGC\,03253 & 104.2 $\pm$  3.5 &   213.2 $\pm$  9.3  & 275.4 $\pm$ 10.6 \tabularnewline 
UGC\,03539 & \,\,\,65.7 $\pm$  3.4 & \,\,\,71.6 $\pm$  3.0 & 142.8 $\pm$ 2.7\,\,\, \tabularnewline 
UGC\,04029 & \,\,\,79.4 $\pm$  5.0 &    \multicolumn{1}{c}{\,\,\,\,\,\,-}   & 174.8 $\pm$ 8.4\,\,\, \tabularnewline 
UGC\,04132 & 100.3 $\pm$  3.1 & 194.9 $\pm$  6.0 & 321.4 $\pm$ 14.6 \tabularnewline 
UGC\,05108 & \,\,\,148.4 $\pm$ 16.0 & 173.5 $\pm$  5.3 & 382.5 $\pm$ 10.8 \tabularnewline 
UGC\,05598 & \,\,\,71.3 $\pm$  2.9 & \,\,\,87.5 $\pm$  2.6 & 187.1 $\pm$ 12.9 \tabularnewline 
UGC\,09542 & \,\,\,72.3 $\pm$  5.2 &    \multicolumn{1}{c}{\,\,\,\,\,\,-}   & 205.6 $\pm$ 6.5\,\,\, \tabularnewline 
UGC\,09873 & \,\,\,79.0 $\pm$  4.0 & \,\,\,64.7 $\pm$  2.6   & 137.5 $\pm$ 5.4\,\,\, \tabularnewline 
UGC\,09892 & \,\,\,66.1 $\pm$  3.6 & \,\,\,79.9 $\pm$  2.3   & 211.7 $\pm$ 9.0\,\,\, \tabularnewline 
UGC\,10123 & \,\,\,88.4 $\pm$  2.4 &    \multicolumn{1}{c}{\,\,\,\,\,\,-}   & 230.2 $\pm$ 15.8 \tabularnewline 
UGC\,10205 & 175.8 $\pm$  2.2 & 143.4 $\pm$ 7.2 & 407.3 $\pm$ 12.3 \tabularnewline 
UGC\,10384 & \,\,\,92.0 $\pm$  3.1 & \,\,\,99.3 $\pm$ 4.7 & 245.2 $\pm$ 16.4 \tabularnewline 
UGC\,10710 & 132.5 $\pm$  5.9 & 100.1 $\pm$  3.5   & 263.7 $\pm$ 21.6 \tabularnewline 

 \hline 
{\bf{ Notes.}} \tabularnewline
 \multicolumn{4}{l}{Column 1: galaxy name; Column 2: median of observed \sigmaR;  Col-} \\
 \multicolumn{4}{l}{umn 3: median of model-based \sigmaR; Column 4: velocity scale.}\\
 \multicolumn{4}{l}{$^a$ model-based \sigmaR and velocity scale of NGC 2730 was calculated } \\
 \multicolumn{4}{l}{ using $l_{\rm \star}$ estimated from $R_{\rm e}$. }
\end{tabular}
%\tablefoot{Sample properties}
\end{table}

\section{DISCUSSION}

\subsection{Uncertainties in \sigmaR}

Sources of uncertainty arise from the calculation of the anisotropy parameters and \sigmaR, these quantities are difficult to determine and require many assumptions \citep[e.g.,][]{hes17,kal17b}.  Recent work has improved our ability to determine these parameters \citep[e.g.,][]{cap08,ger12,ber10,che18,kal17a,mar17,pin18}.  The \sigmaz/\sigmaR\, and \sigmap/\sigmaR\, values we use in this analysis are calculated from parameters derived by \citetalias{kal17b}, who use modern sophisticated modelling to derive them from observations \citep[see ][]{cap08}.

The \sigmaR values we use are derived from \citetalias{fal17}'s CALIFA \sigmao observations.  The data are of high quality but are limited by the spatial resolution, sensitivity and velocity resolution relative to typical \sigmao of the survey, introducing uncertainties to our analysis.  More galaxies and better radial data will improve our characterization of the radial behaviour and help to determine whether the radial trends are a function of other properties.  We apply a dispersion cut-off and $20$\% error cut-off to ensure that we use reliable and accurate data. The dispersion cut-off resulted in many low \sigmaR\, data being excluded from our analysis.  The loss of low quality data points has the largest effect on our analysis at large radii, where there are few high quality data suitable for our analysis.  Despite these uncertainties we can still conclude that \sigmam values overestimate \sigmaR at small $R$ (particularly within $l_{\star}$) and the difference between \sigmam and \sigmaR decreases with increasing $R$ for $R < R_{\rm{e}}$.

 \begin{figure}
 \centering
 \includegraphics[width=0.45\textwidth]{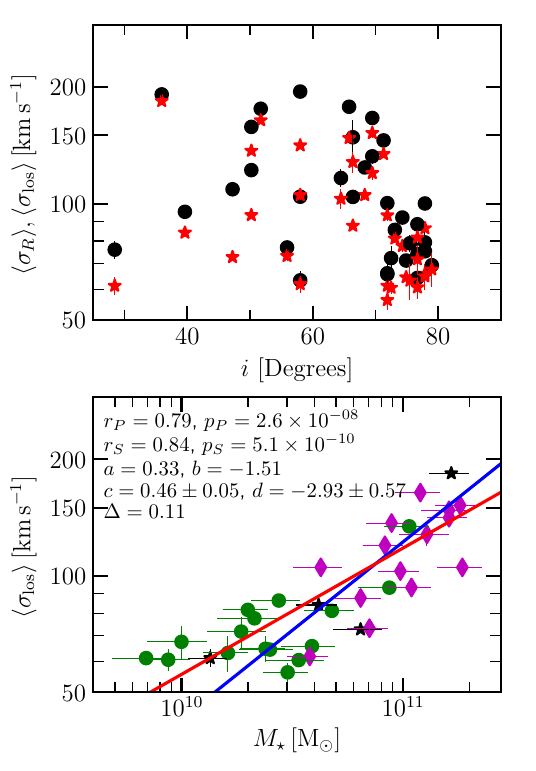}
 \caption{ Top: Plot comparing the radial averages of the radial (black circles) and line-of-sight (red stars) stellar velocity dispersions, averaged over $1\,R_{\rm{e}}$, robustly estimated via the median: \medsigmaR\, and \medsigmalos\, respectively, as a function of inclination. 
 Bottom: \medsigmalos\, as a function of $M_{\star}$.  Galaxies with $i<50^{\circ}$ are shown as black stars, those with $50^{\circ}<i<70^{\circ}$ as magenta diamonds and galaxies with $i>70^{\circ}$ are shown as green circles.  The red line represents the best-fitting line using a robust median-based fit method and the blue line represents the best-fitting line from ODR least-squares fitting.  
 }
 \label{fig:sigmainc}
\end{figure}

 \begin{figure}
 \centering
 \includegraphics[width=0.45\textwidth]{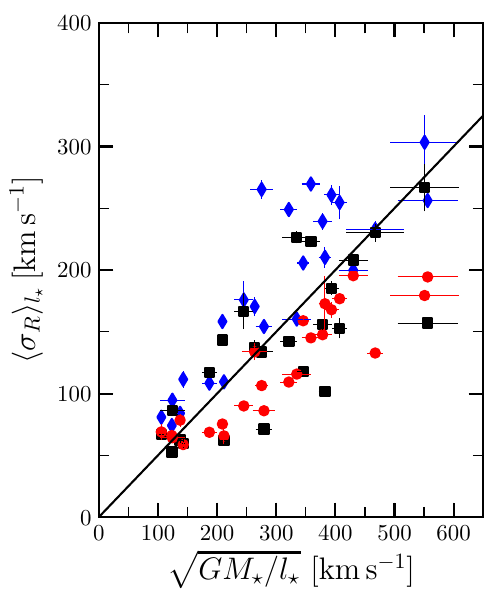}
 \caption{ The radial averages of the stellar velocity dispersion \sigmaRR ( red circles) and model-based velocity dispersions \sigmam$(R)$ all averaged over the stellar scale length $l_{\star}$, robustly estimated via the median, and all plotted against the velocity scale: \gml\,. \sigmam$(R)$ values calculated using $B=0.6$ and a flattening ratio of $7.3$ are shown as blue diamonds and \sigmam$(R)$ calculated using \mbox{\citetalias{kal17b}} $B$ and \mbox{\citet{ber10}} flattening ratios are shown as black squares.  We only show galaxies for which we calculated model-based velocity dispersions.  A \medsigmaR $= 0.5$\gml\, relation is shown by the black line. 
 }
 \label{fig:sigmamodlsv2}
\end{figure}

Inclination has an effect on the observed velocity dispersion because of line-of-sight projection effects and dust extinction.  For highly inclined galaxies, individual fibers cover a wide range of galactocentric radii and galaxy kinematics, therefore each observed spectrum consists of a superposition of a large number of regions with different kinematics.  Variation of the anisotropic stellar velocity ellipsoid complicates the extraction of stellar kinematics parameters further due to the combination of line-of-sight projected velocities and velocity dispersions in the projected spectra. \mbox{\cite{kre05}} also showed how at high inclinations the line-of-sight projection effects cause increased asymmetry in the observed dispersion measurements, resulting in greater differences between the observed and true stellar velocity dispersions.  The increased number of regions covering a wide range of azimuths in line-of-sight observations at high inclination means that Equation 1 becomes a less accurate description of \mbox{\sigmao} in such cases, and its use results in overestimation of \mbox{\sigmao} and hence \sigmaR.  Dust extinction along the line of sight can result in underestimation of the true $R$ of \mbox{\sigmao} measurements, which results in underestimation of \mbox{\sigmao} at low radii.  The interplay between stellar kinematics, inclination and the dust extinction on \mbox{\sigmao} are examined in more detail by \mbox{\cite{kre05}}.  The inclination distribution of our 34 galaxy sample is shown inf Figure \mbox{\ref{fig:sigmainc}}.  The galaxies cover a wide range of inclinations between \mbox{$30^{\circ}$} and \mbox{$80^{\circ}$}, with a large number of galaxies with \mbox{$70^{\circ} < i < 80^{\circ}$}.   

We also look at the relationship between \mbox{\sigmao} and \mbox{$M_{\star}$} and see that the best-fit relationship is similar to the \mbox{\sigmaR} and \mbox{$M_{\star}$} relationship but has slightly weaker correlation and slightly larger rms.  The fitted relations are shown in Figure \mbox{\ref{fig:sigmainc}}, the best-fit ODR relation is: \mbox{$\log \langle\,\sigma_{\rm{obs}}\,\rangle = (0.46 \pm 0.05) \log M_{\star} + (-2.93 \pm 0.57)$} with a root mean squared (rms) of $0.11$.  The figure also shows that galaxies across our inclination range are lie on or close to the best-fit relation.  Some high $i$ and low \mbox{$M_{\star}$} galaxies have either underestimated \mbox{$M_{\star}$} or overestimated \sigmao with respect to the best-fit relationship, both of these can occur due to line-of-sight effects.  We also explore inclination effects as a function of $R$ but find no correlation between \sigmaRR profiles and $i$.  Further investigation and modelling outside of the scope of this paper is required to better constrain the line-of-sight effects on \sigmaR\, and \rm{$M_{\star}$} measurements in the CALIFA sample, but in our analysis, we do not find evidence for $i$ having a strong bias on \sigmaR and its relation with \rm{$M_{\star}$}.

\subsection{Comparison between \sigmaR and \sigmam}

For the comparison with \sigmam we assume $B=0.6$, however Figure \mbox{\ref{fig:dispasym}} shows that typical values of $B$ for our sample are greater than $0.6$. \mbox{\citetalias{kal17b}} also determined flattening ratios for their galaxies in their analysis.  Such analysis can improve \sigmam models but require high quality stellar kinematics data.  We now study the effect of using parameters derived from modelling individual galaxies by determining \sigmam using $B$ and flattening ratios determined by \mbox{\citetalias{kal17b}} and using a relation from \mbox{\citet{ber10}}.  The $B$ values are typically between a factor of one or two greater than the assumed values and the fitted \mbox{\citetalias{kal17b}} on-sky flattening ratios are typically lower by up to a factor of $\sim 2$.  Using these parameters results in small changes in \sigmam that vary between galaxies. However, when we combine the relation that \mbox{\citet{ber10}} fitted between the flattening ratio $q$ and $l_{\star}$: \mbox{$\log(q) = 0.367\log(l_{\star})+0.708$} with \mbox{\citetalias{kal17b}}'s $B$ values to determine \sigmam, we find that the \sigmam values overestimate \sigmaR in most cases but are smaller than those calculated using the parameters we used in the rest of the paper. This is seen in Figure \mbox{\ref{fig:sigmamodlsv2}}, where we plot \mbox{\sigmam}\, radially averaged (calculated using different parameters) over $l_{\star}$ versus the velocity scale. This shows that using better models for $B$ and $q$ can improve \sigmaR predictions, even in the inner regions of galaxies, but still overestimate \sigmaR.  The overestimation is likely due to the departures from non-exponential decline with $R$ of \sigmaR\,, as seen in the varying radial profiles of \sigmaR seen in Figure \mbox{\ref{fig:disps}}. 

The overestimation of \sigmaR has important consequences for stability analysis because lower \sigmaR\, results in lower disc stability.  \mbox{\citet{rom17}} studied the multi-component disc stability, determining the \sigmaR using the \mbox{\citetalias{ler08}} model, and found that inner discs are marginally unstable against non-axisymmetric perturbations and gas dissipation and that the stars drive disc instabilities in the inner regions of galaxies.  Our results indicate that \sigmaR and hence the stability due to stars are overestimated by that model and therefore stars have an even greater effect on disc instabilities than \mbox{\citet{rom17}} found.  The dominance of the stellar disc is contrary to the results of \mbox{\citet{wes14}}, who find that the gas component is more unstable than the stellar component.  Unlike typical studies, they calculate \Sigmas dynamically, resulting in lower \Sigmas than those calculated via population synthesis, as seen when comparing their values to \mbox{\citet{mar13b}} who they draw their sample from.  However, their underestimation may be due to not taking into account the young thin component of the stellar disc and overestimating the scale height \mbox{\citep{ani16}}.  Therefore the uncertainties and assumptions of methods used to determine \Sigmas and $M_{\star}$ should be further investigated to improve $M_{\star}$ estimates.

\subsection{\medsigmaR--$M_{\star}$ relation}

The \medsigmaR--$M_{\star}$ correlation we find is consistent with findings by \citet[][]{bot92}, who found a correlation between \sigmaR and the luminosity of the old disc.  Unlike their luminosity correlation, we find a direct correlation with the stellar mass and this correlation has not been explicitly shown for nearby galaxies in terms of the total stellar mass until this study.  A $\sim 0.5$ power law index would indicate that the \citetalias{ler08} relation: \medsigmaR\,$\sim (\Sigma_{\star}\,l_{\star})^{0.5}$ holds for properties averaged over an effective radius and scale length and is a consequence of discs in hydrostatic equilibrium and isothermal in the vertical direction.  The result of the robust mean fit is a fitted lower power law index of $0.3$, however, this technique does not take into account uncertainties in \sigmaR and $M_{\star}$. Whereas the least-squares ODR fit, which takes into account uncertainties in both parameters, produces a fitted power law index of $0.45 \pm 0.05$.  The constant of proportionality is dependent on the flatness ratio and how close to exponential the discs are, both of which require further analysis and larger samples to better constrain.    

\begin{figure}
 \centering
 \includegraphics[width=0.45\textwidth]{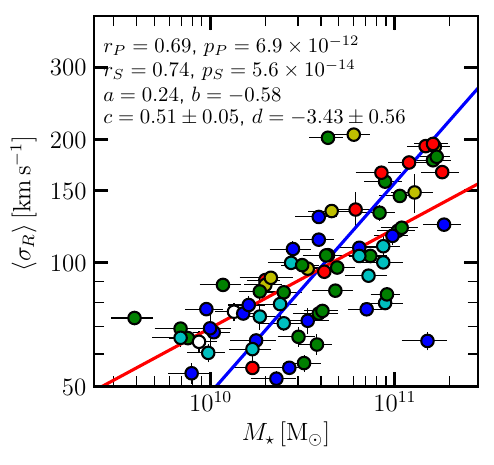}
 \caption{The radial average of the stellar radial velocity dispersion \sigmaRR\, over $1\,R_{\rm{e}}$, robustly estimated via the median,  plotted as a function of $M_{\star}$ for the larger sample of galaxies with \sigmao, circular-speed curve and stellar mass data.  The red line represents the best-fitting line from a robust median-based fit method and the blue line represents the best-fitting line from least-squares fitting.  The galaxies are coded according to Hubble type: Sa galaxies are shown in red, Sab in yellow, Sb in green, Sbc in cyan, Sc in blue and Sd galaxies are shown in white. 
 }
 \label{fig:sigmams}
\end{figure}

Now we test the robustness of the \medsigmaR--$M_{\star}$ relationship against sample size by using a larger sample of spiral galaxies (i.e., Hubble types ranging from Sa to Sd) that have both \sigmaR, circular-speed curves and $M_{\star}$ measurements from \mbox{\citetalias{fal17}}, \mbox{\citetalias{kal17b}} and  \mbox{\citetalias{bol17}}.  This larger sample consists of 74 galaxies.  We plot \medsigmaR\, versus $M_{\star}$ for the sample in \mbox{Figure \ref{fig:sigmams}} and find that the \medsigmaR--$M_{\star}$ correlation still holds.  The significance of the correlation is higher than for the small sample: \mbox{$p_P=6.9 \times 10^{-12}$} and \mbox{$p_S=5.6 \times 10^{-14}$} and the strength of the correlation is $r_P=0.69$ and $r_S=0.74$.  The best-fit parameters are similar to the results for the small sample and the slope of the relationship is closer to $0.5$ when using ODR fits: \mbox{$c=0.51 \pm 0.05$} and \mbox{$d=-3.40 \pm 0.56$}. The rms scatter of the relation is 0.15\,.  The power law index from the robust median fit is $0.24$.  The correlation coefficients and their null hypothesis tests confirm the robustness of the correlation between \medsigmaR\, and $M_{\star}$ regardless of the sample size. 

To test whether the inconsistencies between the ODR and robust median fits is due to uncertainties in the data, we perform least squares fits to the data from the smaller sample, assuming that both parameters have zero uncertainties.  The power law index from this fit is $0.39 \pm 0.05$.  For the larger sample the least-squares fit with zero uncertainties has a power law index fit of $0.34 \pm 0.03$.  Therefore not taking into account the uncertainties of both $M_{\star}$ and \medsigmaR\, results in underestimation of the power law index. When uncertainties are taken into account the power law index of relationship between \medsigmaR\, and $M_{\star}$ is close to $0.5$.

\begin{table}
\caption{
Correlation coefficients and best-fitting parameters for $\delta\log\langle\sigma_{R}\rangle$ versus galaxy properties.}
\label{tab:resid}
\begin{tabular}{lcccc}
\hline 
\multicolumn{1}{l}{Property} & \multicolumn{1}{c}{$r_{\rm{S}}$} & \multicolumn{1}{c}{$p_{\rm{S}}$} & \multicolumn{1}{c}{$c$} & \multicolumn{1}{c}{$d$}  \\
\multicolumn{1}{l}{} & \multicolumn{1}{c}{} & \multicolumn{1}{c}{}  & \multicolumn{1}{c}{} & \multicolumn{1}{c}{}  \\
%\hline
\multicolumn{1}{l}{(1)} & \multicolumn{1}{c}{(2)} & \multicolumn{1}{c}{(3)} & \multicolumn{1}{c}{(4)} & \multicolumn{1}{c}{(5)} \\ 
\hline 

$M_{\star}$ [M$_{\odot}$] & $-0.52$ & \multicolumn{1}{l}{$0.002$} & $-0.49$ $\pm$ 0.15 & \multicolumn{1}{l}{$5.21$ $\pm$ 1.62\,\,\,\,} \tabularnewline 
$M_{\rm{mol}}$ [M$_{\odot}$] & $-0.39$ & $0.022$ & \multicolumn{1}{c}{-} & \multicolumn{1}{c}{-} \tabularnewline 
SFR [\msolyr] & $-0.40$ & $0.018$ & $-0.71$ $\pm$ 0.24 & \multicolumn{1}{l}{$0.24$ $\pm$ 0.10\,\,} \tabularnewline 
 \hline 
 {\bf{Notes.}} \tabularnewline
 \multicolumn{5}{l}{Column 1: galaxy property ; Column 2: Spearman's rank correlation} \\
 \multicolumn{5}{l}{ coefficient; Column 3: $p$-value for Spearman's rank correlation;}\\ 
 \multicolumn{5}{l}{Column 4,5: $c$ and $d$ parameters from the ODR fit $\delta\log\langle\sigma_{R}\rangle$ $=$    } \\
 \multicolumn{5}{l}{$c\log X+d$.} \\

\end{tabular}
\end{table}

To fully characterize the $\langle\sigma_{R}\rangle$--$M_{\star}$ correlation, we have also analysed its scatter:
\begin{equation}
\delta\log\langle\sigma_{R}\rangle=
\log\langle\sigma_{R}\rangle-\log\langle\sigma_{R}\rangle_{\mathrm{fit}}\,,
\end{equation}
where $\log\langle\sigma_{R}\rangle_{\mathrm{fit}}=0.45\,\log M_{\star}-2.78$ is the ODR best-fitting relation (see Table {\ref{tab:corr}}).  The statistical measurements given in Table {\ref{tab:resid}} show that $\delta\log\langle\sigma_{R}\rangle$ has a residual anticorrelation with $M_{\star}$, but this is weaker and less significant than the primary $\langle\sigma_{R}\rangle$--$M_{\star}$ correlation.  This is then a second-order effect, which has no significant impact on our results.

\subsection{\medsigmaR\, relation with other parameters}

The galaxy Main Sequence \citep[e.g.,][]{noe07,dad07,elb07,cat17} shows the correlation between $M_{\star}$\, and SFR.  Hubble type is inversely proportional to stellar mass and metallicity is correlated with stellar mass via the mass-metallicity relation \citep[e.g.,][]{leq79,tre04,san17}.  Therefore the correlation and anticorrelation between SFR, Hubble type and metallicity with \medsigmaR\, can be put in terms of the stellar mass.  \cite{ger12} found a correlation between \sigmaR and molecular gas surface density, therefore the \medsigmaR\,--$M_{\rm mol}$ correlation can be thought of as a reflection of that, and it hints that GMCs may play a role in disc heating. The non-correlation between \sigmaz/\sigmaR\, and \medsigmaR\ is expected \citep[e.g.,][]{ger12} and hints that there is a component of disc heating that only affects \sigmaR.  The \medsigmaR\, versus $M_{\star}$\,, $M_{\rm mol}$\, and SFR relations have similar power law indices, which is consistent with observations that show that the stellar and molecular discs approximately track each other (e.g., \citetalias{bol17}).  

We also study the scatter of the $M_{\rm mol}$\, and SFR relations in a similar manner to $M_{\star}$\, and the correlations and results of the fits are shown in Table \mbox{\ref{tab:resid}}. It should be noted that the applicable \medsigmaRfit\, was used to calculate appropriate $\delta\log\langle\sigma_{R}\rangle$ values for each case, according to the fit results shown in Table \mbox{\ref{tab:corr}}.  The results show that the anticorrelations between $M_{\rm mol}$\, and SFR and their $\delta\log\langle\sigma_{R}\rangle$ are weaker and less significant than for $M_{\star}$\,: $r_S=-0.39$, $p_S=0.022$ for $M_{\rm mol}$\, and $r_S=-0.40$ and $p_S=0.018$ for SFR.  The best-fit relation for $\delta\log\langle\sigma_{R}\rangle$ versus $\log$SFR has a slope of $-0.71$.  We could not achieve a good fit to the data for $\delta\log\langle\sigma_{R}\rangle$ versus $\log M_{\rm mol}$\, using the ODR method.  In both cases the correlations are also much weaker that the fit relations shown in Table \mbox{\ref{tab:corr}}.

\begin{figure*}
 \centering
 \includegraphics[width=\textwidth]{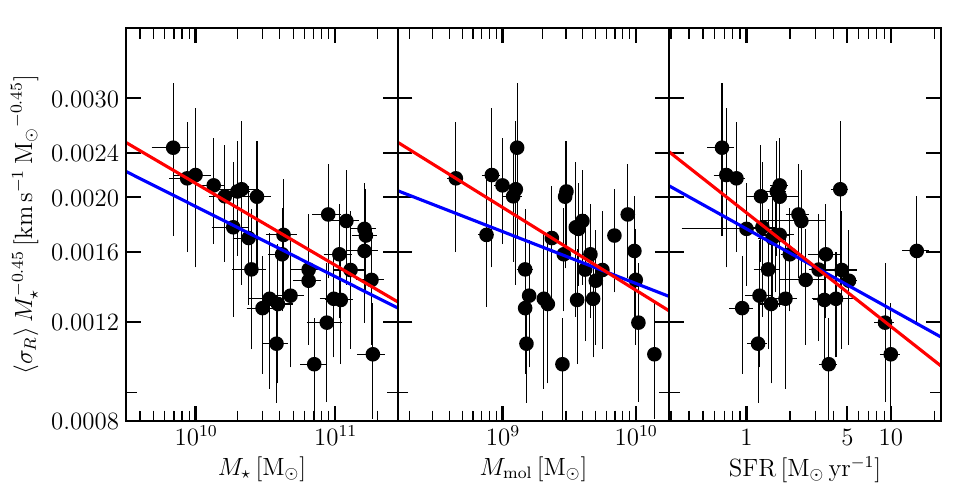}
 \caption{ The radial average of the stellar radial velocity dispersion \sigmaRR over $1\,R_{\rm{e}}$, robustly estimated via the median, divided by $M_{\star}^{0.45}$ and plotted as a function of $M_{\star}$, $M_{\rm{mol}}$ and SFR.  This allows us to study the relationship between the radial average of the stellar radial velocity dispersion and the molecular fraction and specific star formation. The red lines represent the best-fitting lines using a robust median-based fit method and the blue lines represent the best-fitting lines from ODR least-squares fitting.
 }
 \label{fig:meddisppropv2}
\end{figure*}

\begin{table*}
\caption{Correlation coefficients and best-fitting parameters for \sigmaR/$\,M_{\star}^{-0.45}$ versus galaxy properties.}
\label{tab:corr2}
\begin{tabular}{lcccccccccc}
\hline 
\multicolumn{1}{l}{Property} & \multicolumn{1}{c}{$r_{\rm{P}}$} & \multicolumn{1}{c}{$p_{\rm{P}}$} & \multicolumn{1}{c}{$r_{\rm{S}}$} & \multicolumn{1}{c}{$p_{\rm{S}}$} & \multicolumn{1}{c}{$a$} & \multicolumn{1}{c}{$b$}  & \multicolumn{1}{c}{$c$} & \multicolumn{1}{c}{$d$} & \multicolumn{1}{c}{$\Delta$}  \\
\multicolumn{1}{l}{} & \multicolumn{1}{c}{} & \multicolumn{1}{c}{}  & \multicolumn{1}{c}{} & \multicolumn{1}{c}{} & \multicolumn{1}{c}{} & \multicolumn{1}{c}{} & \multicolumn{1}{c}{} & \multicolumn{1}{c}{} & \multicolumn{1}{c}{}  \\
%\hline
\multicolumn{1}{l}{(1)} & \multicolumn{1}{c}{(2)} & \multicolumn{1}{c}{(3)} & \multicolumn{1}{c}{(4)} & \multicolumn{1}{c}{(5)} & \multicolumn{1}{c}{(6)} & \multicolumn{1}{c}{(7)} & \multicolumn{1}{c}{(8)} & \multicolumn{1}{c}{(9)} & \multicolumn{1}{c}{(10)} \\ 
\hline 

$M_{\star}$ [M$_{\odot}$] & $-0.61$ & \multicolumn{1}{l}{$1.5\times10^{-4}$} & $-0.52$ & \multicolumn{1}{l}{$1.7\times10^{-3}$} & $-0.15$ & \multicolumn{1}{l}{\,$-1.22$\,\,\,\,} & $-0.12$ $\pm$ 0.04 & \multicolumn{1}{l}{$-1.47$ $\pm$ 0.40\,\,\,\,} & 0.08 \tabularnewline 
$M_{\rm{mol}}$ [M$_{\odot}$] & $-0.44$ & $9.3\times10^{-3}$ & $-0.39$ & $2.2\times10^{-2}$ & $-0.15$ & \multicolumn{1}{l}{\,$-1.39$\,\,} & $-0.09$ $\pm$ 0.04 & \multicolumn{1}{l}{$-1.93$ $\pm$ 0.42\,\,} & 0.09 \tabularnewline 
SFR [\msolyr] & $-0.47$ & $5.5\times10^{-3}$ & $-0.40$ & $1.8\times10^{-2}$ & $-0.20$ & \multicolumn{1}{l}{\,$-2.73$\,\,} & $-0.14$ $\pm$ 0.05 & \multicolumn{1}{l}{$-2.76$ $\pm$ 0.02\,\,} & 0.09 \tabularnewline 
 \hline 
 {\bf{Notes.}} \tabularnewline
 \multicolumn{10}{l}{Column 1: galaxy property ; Column 2: Pearson's rank correlation coefficient; Column 3: $p$-value for Pearson's rank correlation; Col-} \\
 \multicolumn{10}{l}{umn 4: Spearman's rank correlation coefficient; Column 5: $p$-value for Spearman's  rank correlation; Column 6,7: $a$ and $b$ parameters } \\
 \multicolumn{10}{l}{ from the robust median-based fit $\log \langle \sigma_{R} \rangle$ $= a\log X + b$\,, where X denotes galaxy property; Column 8,9: $c$ and $d$ parameters from the} \\
 \multicolumn{10}{l}{ ODR fit $\log \langle \sigma_{R} \rangle = c \log X + d$; Column 10: rms scatter of scaling relations.   } \\
 \multicolumn{10}{l}{}
\end{tabular}
\end{table*}

We next explore the relationship between \medsigmaR\, and molecular fraction and specific star formation.  We remove the effects of the \medsigmaR\,--$M_{\star}$ correlation and plot \medsigmaR $\,M_{\star}^{-0.45}$ versus $M_{\rm mol}$, $\rm{SFR}$ and $M_{\star}$ for the 34 galaxy sample in \mbox{Figure \ref{fig:meddisppropv2}}.  This allows us to study the aforementioned relationships.  The best-fit relations and correlation coefficients from these plots are shown in \mbox{Table \ref{tab:corr2}}.  \mbox{Figure \ref{fig:meddisppropv2}} shows that there is little correlation between \medsigmaR $\,M_{\star}^{-0.45}$ and $M_{\star}$ for galaxies with $M_{\star}>\,2 \times 10^{10} \rm{M_{\odot}}$ and an anticorrelation between them at smaller $M_{\star}$.  The anticorrelation between $M_{\star}$ and \medsigmaR $\,M_{\star}^{-0.45}$ is weak ($r_S=-0.52$, $p_S=1.7 \times 10^{-3}$) and has a best-fit power law index of $-0.12 \pm 0.04$.  $M_{\rm mol}$ and $\rm{SFR}$ have weaker and less significant anticorrelations with \medsigmaR $\,M_{\star}^{-0.45}$ and their best-fit power law indices range between $-0.09 \pm 0.04$ and $-0.14 \pm 0.05$ respectively.  Their fitted power law indices are consistent with the \medsigmaR $\,M_{\star}^{-0.45}$ versus $M_{\star}$ relation's fitted power law index.  It should be noted that the quantities in the leftmost plot in Figure \ref{fig:meddisppropv2} are directly correlated, and as mentioned before the are also correlations between the quantities in the other plots, therefore care should be taken when attempting to identify correlations from these plots.  Existing $\rm SFR$ and $M_{\rm mol}$ correlations with $M_{\star}$, the consistency between power law indices and the low significance of the correlations (e.g.,\mbox{Table \ref{tab:corr2}}) suggest that the $\rm SFR$ and $M_{\rm mol}$ relationships are dominated by the stronger and more significant $M_{\star}$ anticorrelation that exists at low $M_{\star}$.  However, we require more high quality data to investigate this further and determine whether there are any correlations between \medsigmaR\, and either the molecular fraction or specific star formation.

\section{CONCLUSIONS}

In this study we have used observed line-of-sight \sigmao and fitted dispersion anisotropy parameters to determine \sigmaR for 34 galaxies from the CALIFA survey.  These galaxies cover a wide range of properties such as Hubble types ranging from Sa to Sd.  We compare \sigmaR values to model-based \sigmaR, study how they change with radius and study how they relate to galaxy properties.  Our major conclusions are as follows:

\begin{enumerate}
\item Model-based dispersions overestimate \sigmaR\, at small radii.  The difference can be greater than $50\,$\kph\, within a stellar scale length.  Therefore model-based dispersions do not accurately model \sigmaR and the use of high quality stellar line-of-sight velocity dispersions will result in more accurate stability parameters, asymmetric drift corrections, and better constraints on disc heating processes. 

\item The radial average of \sigmaR over the effective radius is correlated with $M_{\star}$, $M_{\rm mol}$ and SFR, it is weakly correlated with metallicity and weakly anticorrelated with Hubble type.  The \medsigmaR\, versus SFR, metallicity and Hubble type relations can be thought of in terms of the \medsigmaR--$M_{\star}$ relation, which has the strongest and most significant correlation.  And the best-fitting line to the relation is: $\log \langle\,\sigma_{R}\,\rangle = 0.45 \log M_{\star} -2.78$, with a rms scatter of $0.10$ dex compared to 0.12 and 0.18 dex for $M_{\rm mol}$ and SFR using similar samples. For a larger sample of 74 galaxies the best-fitting line to the \medsigmaR--$M_{\star}$ relation is: $\log \langle\,\sigma_{R}\,\rangle = 0.51 \log M_{\star} -3.43$, with an rms scatter of $0.15$ dex.  This \medsigmaR\,$\propto M_{\star}\,^{0.5}$ relation is important and can be used in conjunction with other scaling relations to measure disc stability and to show that nearby disc galaxies self-regulate to a quasi-universal disc stability level \cite[][]{rom18}.

\item The results found in this paper confirm, with a large sample of nearby star-forming spirals, the findings of \citet[][]{rom17}: using observed, rather than model-based, stellar radial velocity dispersions leads to less stable inner galaxy discs and to disc instabilities driven even more by the self-gravity of stars.  This shows, once again, how important it is to rely on high-quality measurements of the stellar line-of-sight velocity dispersion, such as those provided by the CALIFA, SAMI and MaNGA surveys and those promised by second-generation IFU surveys using the Multi Unit Spectroscopic Explorer (MUSE).

\end{enumerate}

\section*{ACKNOWLEDGEMENTS}
KM acknowledges support from the National Research Foundation of South Africa.  We wish to thank the referee for their useful comments and suggestions which helped to improve this paper.  This study uses data provided by the Calar Alto Legacy Integral Field Area (CALIFA) survey (http://califa.caha.es/). Based on observations collected at the Centro Astron\'{o}mico Hispano Alem\'{a}n (CAHA) at Calar Alto, operated jointly by the Max-Planck-Institut f\"{u}r Astronomie and the Instituto de Astrof\'{i}sica de Andaluc\'{i}a (CSIC).

\appendix

%\bsp

\label{lastpage}

\end{document}